\documentclass{article}

\usepackage{arxiv}

\usepackage[utf8]{inputenc} 
\usepackage[T1]{fontenc}    
\usepackage{hyperref}       
\usepackage{url}            
\usepackage{booktabs}       
\usepackage{amsfonts}       
\usepackage{nicefrac}       
\usepackage{microtype}      
\usepackage{lipsum}
\usepackage{graphicx}
\usepackage{caption}
\usepackage{subcaption}
\usepackage[backend=biber, style=authoryear, sorting=nyt]{biblatex}
\usepackage{amsmath}
\graphicspath{ {./images/} }
\title{Approximation of Bayesian Hawkes process with \inlabru}

\author{
 Francesco Serafini \\
  School of Geosciences\\
  University of Edinburgh\\
  \texttt{francesco.serafini@ed.ac.uk} \\
   \And
 Finn Lindgren \\
  School of Mathematics\\
  University of Edinburgh\\
  \texttt{Finn.Lindgren@ed.ac.uk} \\
  \And
 Mark Naylor \\
  School of Geosciences\\
  University of Edinburgh\\
  \texttt{Mark.Naylor@ed.ac.uk} \\
}

\newtheorem{theorem}{Theorem}[section]
\newtheorem{definition}{Definition}[theorem]
\newcommand{\inlabru}[0]{\texttt{inlabru}\,}
\newcommand{\bayesianETAS}[0]{\texttt{bayesianETAS}\, }

\begin{document}
\maketitle
\begin{abstract}
Hawkes process are very popular mathematical tools for modelling phenomena exhibiting a \textit{self-exciting} or \textit{self-correcting} behaviour. Typical examples are earthquakes occurrence, wild-fires, drought, capture-recapture, crime violence, trade exchange, and social network activity. The widespread use of Hawkes process in different fields calls for fast, reproducible, reliable, easy-to-code techniques to implement such models. We offer a technique to perform approximate Bayesian inference of Hawkes process parameters based on the use of the R-package \inlabru. The \inlabru R-package, in turn, relies on the INLA methodology to approximate the posterior of the parameters. Our Hawkes process approximation is based on a decomposition of the log-likelihood in three parts, which are linearly approximated separately. The linear approximation is performed with respect to the mode of the parameters' posterior distribution, which is determined with an iterative gradient-based method. The approximation of the posterior parameters is therefore deterministic, ensuring full reproducibility of the results. The proposed technique only requires the user to provide the functions to calculate the different parts of the decomposed likelihood, which are internally linearly approximated by the R-package \inlabru. We provide a comparison with the \bayesianETAS R-package which is based on an MCMC method. The two techniques provide similar results but our approach requires two to ten times less computational time to converge, depending on the amount of data.     
\end{abstract}


\section{Introduction}
\label{sec:1_intro}

Hawkes processes or \textit{self-exciting} processes, first introduced by \cite{hawkes1971point, hawkes1971spectra}, are counting processes often used to model the "arrivals" of some events over time, when each arrival increases the probability of subsequent arrivals in its proximity. Typical applications can be found in seismology (\cite{ogata1988etas, ogata2006space, ogata2011significant, paik2022nonparametric}), capture-recapture (\cite{altieri2022continuous, weller2018calibration}, invasive species (\cite{balderama2012application}), droughts (\cite{li2021self}), crime (\cite{mohler2011self, mohler2013modeling, mohler2018improving}), finance (\cite{azizpour2018exploring, filimonov2012quantifying, hawkes2018hawkes}), disease mapping (\cite{chiang2022hawkes, garetto2021time}, wildfires (\cite{peng2005space}),  and social network analysis (\cite{kobayashi2016tideh, zhou2013learning}).

Hawkes process, and more in general point processes, are counting processes assuming a value equal to the cumulative number of points recorded in a bounded spatio-temporal region. The main characteristic of a Hawkes process is its ability to model the effect of a point on the probability of observing additional points in its surroundings. For example, in seismology, it is often assumed that each earthquake has the ability to \emph{induce} other earthquakes, and therefore observing an earthquake at a space-time location increases the probability of observing additional earthquakes in its proximity. Therefore, each observed point can be classified as \emph{induced}, if it was induced by another point in the history of the process, or as \emph{background} if it arose spontaneously. In this framework, a Hawkes process can be seen as the superposition of a background process, describing the occurrence of background events, and a sub-process for each observation in the history, describing the occurrence of events induced by that observation. This implies that the rate at which points occur at each space-time location is potentially influenced by the whole history of the process. This makes Hawkes process models non-Markovian. More formal definitions of the Hawkes process, its history, and its conditional intensity are given in Section \ref{sec:2_notation}. 

The application of the Bayesian approach has become increasingly popular also in the Hawkes process field (\cite{rasmussen2013bayesian, donnet2020nonparametric, holbrook2021scalable}). In fact, Hawkes process models are often used in hazard or risk analyses, in which the ability to quantify the uncertainty around quantities of interest (e.g. number of events, probability of events of a certain class, inter-event time distribution) is of paramount importance (\cite{marzocchi2015accounting, smit2019bayesian}). However, applying the Bayesian framework, in these cases, is difficult, given the complex form of the posterior distribution and the high degree of correlation between Hawkes process parameters, and researchers had to resort to frequentist-like estimation techniques (\cite{ebrahimian2014adaptive, omi2015intermediate}). Also, an easy-to-use, extendible, Bayesian technique to handle Hawkes process models is still missing, one of the few examples to the authors' knowledge is represented by \cite{ross2021bayesian}. Furthermore, the techniques habitually used in the literature are based on the Markov-Chain Monte Carlo (MCMC, \cite{robert1999monte}) method which limits the reproducibility of the results and resent from the presence of highly correlated parameters.

In this paper, we propose a novel approximation technique for Hawkes process models based on the use of the Integrated Nested Laplace Approximation (INLA, \cite{rue2017bayesian}) method. The INLA method is a well-known alternative to MCMC methods to perform Bayesian inference. It has been successfully applied in a variety of fields such as seismology (\cite{bayliss2020data}), air pollution (\cite{forlani2020joint}), disease mapping (\cite{riebler2016intuitive, santermans2016spatiotemporal, schrodle2011primer, schrodle2011spatio}), genetics (\cite{opitz2016extensive}), public health (\cite{halonen2015road, }), ecology (\cite{roos2015modeling, teng2022bayesian}), more examples can be found in \cite{bakka2018spatial, blangiardo2013spatial, gomez2020bayesian}. Our approach aims to bring the INLA's advantages to the Hawkes process community and is implemented through the R-package \inlabru. Specifically, the novelty of our approach resides in the likelihood approximation, indeed, the log-likelihood is decomposed in the sum of many small pieces, and each piece is linearly approximated with respect to the posterior mode. This means that the log-likelihood is exact at the posterior mode and the accuracy of the approximation decreases as we move away from that point. Furthermore, the linear approximation and the optimization routine to determine the posterior mode are internally performed by the \inlabru package. The user only has to provide the functions to be approximated, the data, and the priors. The advantages of our approach are both in terms of computational time and simplicity to be extended to include covariates and/or to introduce structure in the parameters (e.g. considering one of them as temporally, or spatially, varying). 

The article is structured as follows: Section \ref{sec:2_notation} introduces the basic definition of a counting process, a Hawkes process, and defines its history and conditional intensity; Section \ref{sec:3_hawkes} describes how Hawkes processes are used in practice and provides some examples on possible choices of the conditional intensity; Section \ref{sec:4_hawkeslike} describes our novel approximation method for the log-likelihood; Section \ref{sec:5_realdata} provides a real data example on the Amatrice seismic sequence and compares the results obtained with our approach with the ones from the \bayesianETAS R-package. For the Amatrice seismic sequence, we also provide a retrospective forecasting experiment in which we predict the daily number of earthquakes; Section \ref{sec:6_sim} shows the results of a simulation experiment in which we simulate the data from a known model and compare the \inlabru and \bayesianETAS implementations. This is done to illustrate how the computational time scales increasing the amount of data. The three appendices at the end of the article (\ref{sec:App.A_post},\ref{sec:App.B_binning}, \ref{sec:App.C_priorvar}) provide the posterior distributions of the parameters for the two implementations considered and perform a sensitivity analysis of the \inlabru results with respect to the binning strategy and the prior choice.

\section{Notation and definitions}
\label{sec:2_notation}

In this section, we give the basic definitions of a counting process, its history, and conditional intensity. Some definitions are only given with respect to time, but they can be easily extended to include space and marking variables. We start with the definition of a counting process. A counting process is a stochastic process assuming integer values changing over time. The value of a counting process at time $t \geq 0$ is equal to the number of observations with time less or equal than $t$. More formally,  

\begin{definition}
A counting process $\{N(t), t \geq 0\}$ is a stochastic process assuming values in the set of non-negative integers $\mathbb N \cup \{0\}$, such that: i) $N(0) = 0$; ii) $N(t)$ is a right-continuous step function with unit increments; iii) $N(T) < \infty$ almost surely if $T < \infty$. Also, given a time interval $[0,T)$ with $T < \infty$, we define the complete set of observations up to time $T$ as $\mathcal H_T = \{t_h: t_h \in [0, T) \,\forall h = 1,....,N(T^{-})\}$. Given a random $t \in [0,T)$ we define the \emph{history} of the process up to time $t$ as the subset of elements of $\mathcal H_T$ recorded \emph{strictly} before $t$ and we call it $\mathcal H_t = \{t_h \in \mathcal H_T : t_h < t\}$.
\label{def:2.0.1}
\end{definition} 

Definition \ref{def:2.0.1} can be extended to the marked spatio-temporal case. In this case, a generic observed point is $\mathbf x = (t, \mathbf s, m)$ and is composed of a time $t$, a spatial location $\mathbf s$, and a marking variable $m$. The domain is given by $\mathcal X = [0,T) \times W \times M$, where $T > 0$, $W \subset \mathbb R^2$ and $M \subseteq \mathbb R$. The value of the counting process at time $t$ is the number of events recorded before $t$ (included), with spatial location in $W$ and marking variable in $M$. Assuming that the spatial region of interest ($W$) and the marking variable's domain ($M$) are constant over time, we can use the same notation for the complete set of observations and the history of the process.
In this case, the complete set of observations is $\mathcal H_T = \{\mathbf x_h = (t_h, \mathbf s_h, m_h) : \mathbf x_h \in \mathcal X \, \forall h = 1,...,N(T^{-}) \}$, and the history of the process becomes $\mathcal H_t = \{\mathbf x_h = (t_h, \mathbf s_h, m_h) \in \mathcal H_T: t_h < t\}$. 

Any counting process can be defined by specifying its conditional intensity. The conditional intensity of a counting process at time $t$ is the expected infinitesimal rate at which events occur around time $t$ given the history of the process $\mathcal H_t$. More formally,

\begin{definition}
For a counting process $\{N(t), t \geq 0\}$ with history $\mathcal H_t$, the conditional intensity function of the process $N(t)$ is:

\begin{equation}
    \lambda(t | \mathcal H_t) = \lim_{\Delta_t \downarrow 0} \frac{\mathbb E[N(t + \Delta_t) - N(t^{-})
    \mid \mathcal H_t]}{\Delta_t} \nonumber
\end{equation}

For $\Delta_t, t \geq 0$. Assuming that the limit exists, the conditional intensity is left-continuous and $\lambda(t|\mathcal H_t) \geq 0, \, \forall t \geq 0$.
\label{def:2.0.2}
\end{definition}

Definition \ref{def:2.0.2} can also be extended to include a space location and a marking variable. The conditional intensity $\lambda(\mathbf x | \mathcal H_t)$ is the expected infinitesimal rate at which points occur in $(t, t + \Delta_t), \Delta_t > 0$, around space location $\mathbf s$, with marking variable around $m$. 

The first characteristic for a Hawkes process as defined in \cite{hawkes1971spectra} Equation (4) is that the probability of the number of events in $(t, t + \Delta_t)$ being equal to $n = 0,1,...$ is given by:

\begin{equation}
    \Pr ( N(t + \Delta_t) - N(t) = n \vert \mathcal H_t) = 
    \begin{cases}
    1 - \lambda(t)\Delta_t - o(\Delta_t) & if \, n = 0 \\
    \lambda(t) \Delta_t + o(\Delta_t) & if \, n = 1 \\
    o(\Delta_t) & if \, n > 1
    \end{cases}
    \label{eq:1_conditions}
\end{equation}

Equation \ref{eq:1_conditions} has two major implications. The first one is that the probability of having more than one event in an infinitesimal interval around $t$ goes to zero faster than the length of the interval. This implies that the probability of observing two events at the same time is zero and that the number of events in $\mathcal H_T$ is equal to $N(T)$ with probability one. However, recorded data doesn't have to obey that (due to time discretisation). The second is that the probability of having an event in $(t, t + \Delta_t)$ conditional on the history $\mathcal H_t$, for small $\Delta_t > 0$, is completely specified by the conditional intensity. 

Now, we can define a Hawkes process model through its conditional intensity:

\begin{definition}
A Hawkes process is a counting process with conditional intensity given by:
\begin{equation}
    \lambda(\mathbf x | \mathcal H_t) = \mu(\mathbf x) + \sum_{\mathbf x_h \in \mathcal H_t} g(\mathbf x, \mathbf x_h)
    \label{eq:2_hawkes_def}
\end{equation}
Where $\mu: \mathcal X \rightarrow [0,\infty)$, and $g:\mathcal X \times \mathcal X \rightarrow [0,\infty)$
\label{def:2.0.3}
\end{definition} 

The conditional intensity is composed of a part $\mu(\mathbf x)$ usually called the background rate, which does not depend on the history; and a second part representing the contribution to the intensity from the points in the history. The function $g:\mathcal X \times \mathcal X \rightarrow \mathbb R^{+}$ is known as \textit{excitation} or \textit{triggering} function and measures the influence of observation $\mathbf x_h$ on the point $\mathbf x$. 

Definition \ref{def:2.0.3} implies that the whole history of the process is important to determine the current level of intensity. In this view, Hawkes processes can be seen as a non-Markovian extension of inhomogeneous Poisson processes. Both the background rate and the triggering function depends on a set of parameters $\boldsymbol \theta \in \Theta \subset \mathbb R^m$ which determines the properties of the Hawkes process under study (e.g. number of events per time interval, probability of a certain type of events, average number of induced events, type of clustering). Our technique provides a way to have a fully-Bayesian analysis of the parameters $\boldsymbol \theta$.                       

\section{Hawkes process modelling}
\label{sec:3_hawkes}

The Hawkes process intensity in Equation \ref{eq:2_hawkes_def} is composed by two part, a background rate $\mu(\mathbf x)$ and an \textit{excitation} or \textit{triggering} function $g(\mathbf x, \mathbf x_h)$. The background rate and the triggering function depend upon a number of parameters $\boldsymbol \theta$. Our objective is to provide a technique to determine the posterior distribution of $\boldsymbol \theta$ having observed points in $\mathcal X = [0,T] \times W \times M$. Equation \ref{eq:2_hawkes_def} also shows that a Hawkes process can be thought of as the sum of $n + 1$ Poisson processes, where $n = N(T)$ is the number of observations in the history of the process up to time $T<\infty$. One Poisson process represents the background rate and has intensity $\mu(\mathbf x)$, the others $n$ Poisson processes are each one generated by an observation $\mathbf x_h$ and have intensity $g(\mathbf x, \mathbf x_h)$. Many algorithms for fitting Hawkes process models are based on this decomposition and make use of a latent variable assigning the points to one of those $n + 1$ Poisson processes (\cite{ross2021bayesian, veen2008estimation}). Our approach is different because there is no explicit or implicit classification of the points into background and induced events. 

Regarding marked spatio-temporal Hawkes process models, we only report the case where the marking variable distribution is independent of space and time, we refer to this distribution with $\pi(m)$. For the case where this assumption does not hold, and we have $\pi(\mathbf x = (t, \mathbf s, m))$, we just need to substitute $\mu(\mathbf x)$, and $g(\mathbf x, \mathbf x_h)$ with $\mu(\mathbf x)\pi(\mathbf x)$, and $g(\mathbf x, \mathbf x_h)\pi(\mathbf x)$ in all the following expressions without loss of generality. This is valid for both discrete and continuous distribution of the marking variable. Assuming an independent marking variable distribution the Hawkes process conditional intensity is given by:

\begin{equation}
    \lambda(\mathbf x = (t,\mathbf s, m)| \mathcal H_t) = \left(\mu(\mathbf x) + \sum_{\mathbf x_h \in \mathcal H_t} g(\mathbf x, \mathbf x_h)\right)\pi(m)  
    \label{eq:3_mrkd.space.time.def}
\end{equation}

Given the assumption of independence between the process representing the space-time locations and the marking variable's distribution, we only focus on the distribution of the space-time locations. The parameters of the marking variable distribution will be estimated independently and based on the observed marks solely. This is the usual situation in seismology, where the marking variable is the magnitude of the event, and its distribution is usually assumed to be independent of the space-time location of the events. If the assumption does not hold, applying the substitution described above allows us to estimate the marking variable distribution's parameters along with the Hawkes process parameters.

In this paper, we consider a spatially varying background rate that remains constant over time. This is done mainly to limit the number of modes in the likelihood and the correlation between parameters. Furthermore, we are going to consider a background rate parameterized as

\begin{equation}
    \mu(\mathbf x) = \mu u(\mathbf s)
    \label{eq:4_bkgdef}
\end{equation}

with $\mu \geq 0$ representing the number of expected background events in the area for a unit time interval, and $u(\mathbf s)$ represents the spatial variation of the background rate and we assume it is normalized to integrate to one over the spatial domain. Different techniques have been employed to estimate $u(\mathbf s)$. For example, in seismology, it is common practice to estimate it independently from the parameters of the triggering function smoothing a declustered set of observations (\cite{ogata2011significant}).   

The common approach to model the triggering function is to factorize it in different components representing the effect of the observations $\mathbf x_h$ on the evaluation point $\mathbf x$ on the different dimensions (i.e. time, space, marking variable). More formally,

\begin{equation}
    g(\mathbf x, \mathbf x_h) = g_m(m_h)g_t(t - t_h)g_{\mathbf s}(\mathbf s - \mathbf s_h)\mathbb I(t > t_h)
    \label{eq:5_trigdec}
\end{equation}

Where, $I(t > t_h)$ is an indicator function assuming value one when the condition holds, and zero otherwise. The function $g_m(m_h)$ is the marking variable triggering function representing the effect of different values of the marking variable (e.g. if $m$ is the magnitude of an earthquake, large earthquakes have a stronger influence); $g_t(t - t_h)$ is the time triggering function determining the time decay of the observed point's effect, and it is usually a decreasing function of $t - t_h$; $g_\mathbf s(\mathbf s - \mathbf s_h)$ is the space triggering function which has the same role of the time triggering function but in space and is usually a function of the \textit{distance} between points (different distances may be employed). 

Following this decomposition, also the parameter vector $\boldsymbol \theta$ can be decomposed in $\boldsymbol \theta = (\boldsymbol \theta^{(\mu)}, \boldsymbol \theta^{(m)}, \boldsymbol \theta^{(t)}, \boldsymbol \theta^{(\mathbf s)})$, where $\boldsymbol \theta^{(\mu)}$ represents the parameters of the background rate, and $\boldsymbol \theta^{(m)}$, $\boldsymbol \theta^{(t)}$, $\boldsymbol \theta^{(\mathbf s)}$ represent, respectively, the parameters of the magnitude, time and space triggering functions. We call $J_\mu, J_m, J_t, J_\mathbf s$ the set of indexes indicating, respectively, the position of the background rate, marking variable triggering function, time triggering function, and space triggering function parameters inside $\boldsymbol\theta$, so we can write $\boldsymbol \theta_\mu = \{\theta_j \in \boldsymbol{\theta}: j \in J_\mu\}$. This notation will be particularly useful in Section \ref{sec:4_hawkeslike}.  

Table \ref{tab:1_trig.funs} reports some of the typical choices for the space-time triggering function. Many modifications of these functions are used in real-data applications. For example, we can imagine a different time or space effect for different values of the marking variable. In seismology, it is common to consider a magnitude-dependent space triggering function representing the fact that earthquakes with large magnitudes affect wider areas. Another modification usually found in applications is to consider the normalized version of the reported functions to ensure they integrate to one over the (respective) domain. 

\begin{table}
 \caption{Typical choices of time and space triggering functions}
  \centering
  \begin{tabular}{lll}
    \toprule
    \multicolumn{2}{c}{Time triggering}  \\                 \\
    Name     & function   & parameters  \\
    \midrule
    Exponential & $\beta e^{-\alpha(t - t_h)}$ & $\alpha, \beta \geq 0$       \\ \\
    Power Law  & $k \left( 1 + \frac{t - t_h}{c}\right)^{-p}$ & $k \geq 0, c > 0, p > 1$ \\ \\ 
    \midrule
    \multicolumn{2}{c}{Space triggering}     \\  \\
    Gaussian & $\det(2\pi\Sigma)^{-1/2} e^ {-\frac{1}{2} (\mathbf s - \mathbf s_h)^T \Sigma^{-1} (\mathbf s - \mathbf s_h) } $   & $\Sigma $ positive semi-definite   \\ \\ 
    Power Law & $(1 + \frac{d(\mathbf s, \mathbf s_h)}{\gamma})^{-q}$ & $\gamma > 0, q > 1$ \\ 
    \bottomrule
  \end{tabular}
  \label{tab:1_trig.funs}
\end{table}

As explained in \cite{laubelements}, the choice of the triggering function is crucial to the reliability and stability of any estimation procedure for Hawkes process parameters. For example, many techniques use triggering functions normalized to integrate to 1 over an infinite domain. For the approximation illustrated in this paper, we recommend using functions as close to linearity as possible with respect to the parameters, and for the author's experience, the unnormalized version works best. The motivations behind this requirement will be illustrated in the next section. 
                
In the real data example provided in Section \ref{sec:5_realdata}, we apply our technique to earthquake data. The data is supposed to come from a spatio-temporal marked Hawkes process model, where the marking variable is the magnitude, however, we will consider it as a temporal marked point process, ignoring the information on the spatial location. The effect of that is to replace the full space-time intensity with a spatially integrated intensity. Indeed, assuming that the region of interest is constant over time, any temporal model, with intensity $\lambda'$ can be seen as a spatio-temporal model (with intensity $\lambda$) integrated over space,

\begin{equation}
    \lambda'(t,m | \mathcal H_t) = \int_W \lambda(t,\mathbf s, m | \mathcal H_t)d\mathbf{s}
\end{equation}

where $W \subset \mathbb R^2$. For the spatio-temporal model, if the background rate is given by equation \ref{eq:4_bkgdef} and the triggering function by equation \ref{eq:5_trigdec}, the temporal background rate ($\mu'$) and triggering function ($g'_t$) are given by

\begin{align}
    \mu' & = \mu \int_{W} u(\mathbf s) d\mathbf s \\ g'_t(t - t_h) & = g_t(t-t_h)\int_{W} g(\mathbf s - \mathbf{s_h})d\mathbf s
\end{align}

Regarding the background rate, if $u(\mathbf s)$ is normalized to integrate to 1 over the domain, the background rate is the same as in the spatio-temporal. For the triggering function, if there were no boundary effects, the integral would be independent of $\mathbf s_h$, so it would just be a common amplitude scaling. This seems a reasonable simplification to be able to treat space-time data as temporal only.

\section{Hawkes process log-likelihood approximation}
\label{sec:4_hawkeslike}

In this section, we illustrate our Hawkes process log-likelihood approximation technique. This approximation technique is new and allows us to express the Hawkes process log-likelihood as a sum of linear functions of the parameters $\boldsymbol{\theta}$. Suppose to have observed $n$ events $\mathcal H_{T_1, T_2} = \{\mathbf x_1,..., \mathbf x_n : \mathbf x_i \in \mathcal X \, \forall i = 1,...,n\}$, where $\mathcal X = [T_1, T_2] \times W \times M$, with $0 \leq T_1 < T_2 < \infty$, $W \subset \mathbb R^2$, and $M \subseteq \mathbb R$. To ease the notation in the next steps we are using $\mathcal H = \mathcal H_{T_1, T_2}$ to indicate the complete set of observations. The general point process model log-likelihood given the observations is:

\begin{equation}
    \mathcal L(\boldsymbol \theta | \mathcal H) = - \Lambda(\mathcal X | \mathcal H) + \sum_{h = 1}^n \log\lambda(\mathbf x_h | \mathcal H_{t_h})
    \label{eq:6_pploglik}
\end{equation}

where $\mathcal H_{t_h}$ is the subset of $\mathcal H_{T_1, T_2}$ of events recorded strictly before $t_h$ and,

\begin{equation}
    \Lambda(\mathcal X | \mathcal H) = \int_{\mathcal X} \lambda(\mathbf x | \mathcal H) d\mathbf x
    \label{eq:7_Lambdadef}
\end{equation}

is the integrated conditional intensity corresponding to the expected number of points in $\mathcal X$. The integrated conditional intensity can be decomposed using the branching structure of Hawkes processes, indeed, we can think of the expected number of points in an area as the expected number of background points plus the expected number of points induced by each observation in the history. Formally, having observed $n = | \mathcal H_{T_1, T_2} | $ events,

\begin{equation}
    \Lambda(\mathcal X | \mathcal H) = \Lambda_0(\mathcal X) + \sum_{h = 1}^n \Lambda_h(\mathcal X) 
    \label{eq:8_Lambdadec}
\end{equation}

where,

\begin{equation}
    \Lambda_0(\mathcal X) = \int_{\mathcal X} \mu(\mathbf x) d\mathbf x = (T_2 - T_1)\mu 
    \label{eq:9_Lambda0}
\end{equation}

is the integrated background rate, and is interpreted as the number of expected background events. The last equation only holds if the background rate follows the definition in Equation \ref{eq:4_bkgdef}. The other quantity is given by

\begin{equation}
    \Lambda_h(\mathcal X) = \int_{\mathcal X} g(\mathbf x, \mathbf x_h) d\mathbf x = g_m(m_h)\int_{\max(T_1, t_h)}^{T_2} \int_W g_t(t - t_h)g_{\mathbf s}(\mathbf s - \mathbf s_h) dt d\mathbf s 
    \label{eq:10_Lambdah}
\end{equation}

and is interpreted as the number of expected points generated by the observation $\mathbf x_h$. The last equation only holds if we use Equation \ref{eq:5_trigdec} to define the triggering function.

The log-likelihood can be decomposed into three main components:

\begin{equation}
    \mathcal L(\boldsymbol \theta) = - \Lambda_0(\mathcal X) - \sum_{h = 1}^n \Lambda_h(\mathcal X) + \textrm{SL}(\mathcal H)
    \label{eq:11_loglikdec}
\end{equation}

The expected number of background events $\Lambda_0(\mathcal X)$, the expected number of induced events $\sum_h \Lambda_h(\mathcal X)$, and the sum of the log-intensities $\textrm{SL}(\mathcal H) = \sum_h \log\lambda(\mathbf x_h|\mathcal H_{t_h})$. 

Our technique is based on approximating these three components separately. The approximation is such that the value of the log-likelihood is exact at the posterior mode $\boldsymbol \theta^*$, and the degree of accuracy decays as we move from there. The level of accuracy for values of the parameters far from the posterior mode strongly depends on the choice of the triggering functions. Specifically, we separately perform a linear approximation of $\log\Lambda_0(\mathcal X)$, $\log\Lambda_h(\mathcal X)$, and $\log\lambda(\mathbf x_h)$, for $h = 1,...,n$, and therefore, these functions should be as close to being linear as possible.   

The next subsections illustrate the approximation of the different log-likelihood components. The last subsection reports some details on the iterative algorithm used to determine the mode of the posterior distribution around which the approximation is performed. For all of them, we will make explicit the dependence of the log-likelihood components from $\boldsymbol \theta$ and omit dependence from the domain $\mathcal X$, formally, $\Lambda(\mathcal X) = \Lambda(\mathcal X, \boldsymbol \theta) = \Lambda(\boldsymbol \theta)$. Also, if a quantity is approximated we use the Tilde symbol, such that $\Tilde{f}(x)$ is the approximation of $f(x)$, while over-lined quantities stand for linearised, such that $\overline{f}(x, x_0)$ is the linear version of $f(x)$ with respect to $x_0$.

\subsection{Part I - Expected Number of background events}
\label{sec:4.1_part1}

We approximate the integrated background rate using a linear approximation of its logarithm. Namely, 

\begin{equation}
    \Tilde{\Lambda}_0(\boldsymbol{\theta}) = \exp\{\overline{\log \Lambda}_0( \boldsymbol \theta, \boldsymbol \theta^*)\} 
    \label{eq:12_L0approx}
\end{equation}

where,

\begin{equation}
    \overline{\log\Lambda}_0(\boldsymbol \theta, \boldsymbol \theta^*) = \log\Lambda_0(\boldsymbol \theta^*) + \frac{1}{\Lambda_0(\boldsymbol \theta^*)}\sum_{j = 1}^m (\theta_j - \theta^*_j) \frac{\partial}{\partial \theta_j} \Lambda_0(\boldsymbol \theta) \Bigg |_{\boldsymbol \theta = \boldsymbol \theta^*}
    \label{eq:13_L0linearapprox}
\end{equation}

This approach is particularly convenient if the background rate has the form reported by Equation \ref{eq:4_bkgdef}. The only parameter to estimate using this approximation is $\mu \geq 0$. Changing parameter to $\theta_{\mu} = \log \mu$, we have two huge advantages. First, $\theta_{\mu} \in (-\infty, \infty)$ is a free-constraint parameter, and second, the logarithm of the expected number of background events is linear in $\theta_{\mu}$, which means that there will be no approximation at this step and this component will be exact for any value of $\theta_\mu$.

\subsection{Part II - Expected Number of triggered events}
\label{sec:4.2_part2}

We start the approximation of the expected number of triggered events by considering the expected number of events triggered by a single observation $\mathbf x_h$. This is given by Equation \ref{eq:10_Lambdah}. Considering a partition of the space $\mathcal X$, namely $b_{1,h},...,b_{B_h,h}$ such that $\bigcup_i b_{i,h} = \mathcal X$ and $b_{j,h} \bigcap b_{i,h} = \emptyset,\, \forall i \neq j$, we can write:

\begin{equation}
    \Lambda_h(\boldsymbol \theta) = \sum_{i = 1}^{B_h} \int_{b_{i,h}} g(\mathbf x, \mathbf x_h) d\mathbf x = \sum_{i = 1}^{B_h} \Lambda_h(b_{i,h}, \boldsymbol \theta)
    \label{eq:14_Lhapprox}
\end{equation}

We approximate the above quantity linearly approximating the logarithm of the elements of the summation. This increase the computational time and memory required by the algorithm but it provides a much better approximation than considering one bin only. More formally, 

\begin{equation}
    \Tilde{\Lambda}_h(\boldsymbol \theta) = \sum_{i = 1}^{B_h} \exp\{ \overline{\log\Lambda}_h(b_{i,h}, \boldsymbol \theta, \boldsymbol \theta^*)\}
    \label{eq:15_Lhdecomp}
\end{equation}

Where $\overline{\log\Lambda}_h(b_{i,h}, \boldsymbol \theta, \boldsymbol \theta^*)$ is the linear approximation with respect to the posterior mode of the expected number of generated events by the observation $\mathbf x_h$ in the area $b_{i,h}$ and has the same form of Equation \ref{eq:13_L0linearapprox}. 

Assuming that we are dealing with a spatio-temporal marked Hawkes process model with triggering function given by Equation \ref{eq:5_trigdec} and bins partitioning the time domain only, such that $b_{i,h} = [t_{i-1, h}, t_{i,h}) \times W$ for $i = 1,...,B_h$ and $t_{i,h} < t_{j,h} \forall i < j$ and $t_0 = \max(T_1, t_h)$ and $t_B = T_2$, we have that:

\begin{align}
    \Lambda_h(b_{i,h}, \boldsymbol \theta) & = g_m(\mathbf x_{t_h}, \boldsymbol \theta^{(m)})\left(\int_{t_{i-1,h}}^{t_{i,h}} g_t(t - t_h, \boldsymbol \theta^{(t)}) dt\right) \left(\int_W g_{\mathbf s}(\mathbf s - \mathbf s_h, \boldsymbol \theta^{(\mathbf s)}) d\mathbf s\right) \nonumber \\ 
    & = g_m(m_h, \boldsymbol \theta^{(m)}) I_t(b_{i,h}, \boldsymbol \theta^{(t)}) I_{\mathbf s}(\boldsymbol \theta^{(\mathbf s)})
    \label{eq:16_Lhspacetime}
\end{align}

where $I_t(b_{i,h},\boldsymbol \theta^{(t)})$ and $I_{\mathbf s}(\boldsymbol \theta^{(\mathbf s)})$ are, respectively, the integral of the time and space triggering function. The derivative of the logarithm of $\Lambda_h(b_{i,h}, \boldsymbol \theta)$ with respect to $\theta_j \in \boldsymbol \theta$ is given by

\begin{equation}
    \frac{\partial}{\partial \theta_j} \log \Lambda_h(b_{i,h}) = 
    \begin{cases}
    \frac{\partial}{\partial \theta_j} \log g_m(m_h), & \text{if $j \in J_m $} \\ 
    \frac{\partial}{\partial \theta_j} \log I_t , & \text{if $j \in J_t $} \\ 
    \frac{\partial}{\partial \theta_j} \log I_\mathbf s , & \text{if $j \in J_\mathbf s $}
    \end{cases}
    \label{eq:17_logLhder}
\end{equation}

Where $J_m, J_t, J_{\mathbf s}$ are defined in Section \ref{sec:3_hawkes}.

Therefore, the accuracy of the approximation depends on \textit{how close} to be linear the functions $\log g_m(\cdot), \log I_t(\cdot), \log I_{\mathbf s}(\cdot)$ are with respect the parameters $\boldsymbol \theta$. In the case of normalized triggering functions, we have $\Lambda_h(\mathcal X) = g_m(m_h)$. This means that, on one hand, we don't need to split the integral in different bins saving computational time and memory; on the other hand, the information on the parameters $\theta_j \in \boldsymbol \theta^{(t)} \bigcup \boldsymbol \theta^{(\mathbf s)}$ provided by this likelihood component is lost. Also, normalized triggering functions tend to be \textit{farther} from linearity than the corresponding unnormalized versions and this is crucial for the approximation of the sum of log-intensities.

We remark that the division in bins is essential for the accuracy of the approximation and the ability to converge of the algorithm. Different binning strategies can be employed, and their performance depends on the form of the triggering function. For example, in the case in which the time triggering function represents the time-decay of the influence of an observation on the intensity, we expect it to be a monotonic decreasing function of the time difference and, therefore, a convenient strategy would be to consider a denser partition around zero and larger bins far from it where the function flattens. In Appendix \ref{sec:App.B_binning} we illustrate the binning strategy used in the real data and simulation examples which has the characteristics described above. In there, we perform a sensitivity analysis fitting the same Hawkes process model using different binning strategies, and Table \ref{tab:5_bins} compares the different binning strategies in terms of computational time and ability to converge.

\subsection{Part III - Sum of log-intensities}
\label{sec:4.3_part3}

For the sum of log-intensities calculated at the observed points, we simply consider the linear approximation of the elements of the summation, namely 

\begin{equation}
    \Tilde{\textrm{SL}}(\mathcal H) = \sum_{h = 1}^n \overline{\log\lambda}(\mathbf x_h, \boldsymbol \theta, \boldsymbol\theta^*)
    \label{eq:18_SLapprox}
\end{equation}

where, omitting the dependence from $\mathbf x_{h}$,

\begin{equation}
    \overline{\log\lambda}(\mathbf x_h, \boldsymbol \theta, \boldsymbol \theta^*) = \log\lambda(\boldsymbol \theta^*) + \frac{1}{\lambda(\boldsymbol \theta^*)}\sum_{j = 1}^m (\theta_j - \theta^*_j) \frac{\partial}{\partial \theta_j} \lambda(\boldsymbol \theta) \Bigg |_{\boldsymbol \theta = \boldsymbol \theta^*}
    \label{eq:19_SLlinearapprox}
\end{equation}

which is the same as Equation \ref{eq:13_L0linearapprox}. 

Assuming to be interested in a spatio-temporal marked Hawkes process model, with background rate specified by Equation \ref{eq:4_bkgdef}, considering $u(\mathbf s)$ known for any $\mathbf s \in W$, and triggering function specified by Equation \ref{eq:5_trigdec}, the conditional intensity is given by:

\begin{equation}
    \lambda(\mathbf x_h | \mathcal H_{t_h}) = \mu u(\mathbf s_h) + \sum_{k : \mathbf x_k \in \mathcal H_{t_h}} g_m(m_k)g_t(t_h - t_k)g_{\mathbf s}(\mathbf s_h - \mathbf s_k) 
    \label{eq:20_lambda_spacetime}
\end{equation}

with derivative with respect to $\boldsymbol \theta$ equal to 

\begin{equation}
     \frac{\partial}{\partial \theta_j} \lambda(\mathbf x_h) =  
    \begin{cases}
     u(\mathbf s_h), & \text{if $\theta_j =  \mu$} \\ \\
    \sum_k g_t(t_h-t_k)g_{\mathbf s}(\mathbf s_h - \mathbf s_k) \frac{\partial}{\partial \theta_j} g_m(m_k) , & \text{if $j \in  J_m$} \\ \\ 
    \sum_k g_m(m_k)g_{\mathbf s}(\mathbf s_h - \mathbf s_k) \frac{\partial}{\partial \theta_j} g_t(t_h - t_k) , & \text{if $j \in  J_t$} \\ \\
    \sum_k g_m(m_k)g_t(t-t_k)\frac{\partial}{\partial \theta_j} g_{\mathbf s}(\mathbf s_h - \mathbf s_k) , \quad & \text{if $j \in  J_\mathbf s$}
    \end{cases}
    \label{eq:21_lambdader}
\end{equation}

The above expression indicates that the accuracy of the approximation depends on how close to linearity the different triggering function components are.

\subsection{Full approximation and \inlabru implementation}
\label{sec:4.4_fullapprox}

Putting all together, the Hawkes process log-likelihood approximation used by our technique is: 

\begin{align}
    \Tilde{\mathcal L}(\boldsymbol \theta, \boldsymbol \theta^*) & = - \Tilde{\Lambda}_0(\boldsymbol \theta, \boldsymbol \theta^*) - \sum_{h = 1}^n\sum_{i=1}^{B_h}\Tilde{\Lambda}_h(b_{i,h},\boldsymbol \theta, \boldsymbol \theta^*) + \Tilde{SL}(\mathcal H, \boldsymbol \theta, \boldsymbol \theta^*) \nonumber\\ 
    & = -\exp\{\overline{\log \Lambda}_0( \boldsymbol \theta, \boldsymbol \theta^*)\} - \sum_{h = 1}^n \sum_{i = 1}^{B,h} \exp\{ \overline{\log\Lambda}_h(b_{i,h}, \boldsymbol \theta, \boldsymbol \theta^*)\} + \sum_{h = 1}^n \overline{\log\lambda}(\mathbf x_h, \boldsymbol \theta, \boldsymbol\theta^*)
    \label{eq:22_loglikapprox}
\end{align}

The approximation is performed with respect to the mode of the posterior distribution $\boldsymbol \theta^*$, which is determined by an iterative algorithm. The algorithm starts from a linearisation point $\boldsymbol \theta^*_0$ (provided by the user), finds the mode of the linearised (with respect to $\boldsymbol \theta^*_0$) posterior using the INLA method, namely $\overline{\boldsymbol\theta}^*_1$, the value of the linearisation point is updated to $\boldsymbol \theta^*_1 = \alpha\boldsymbol{\theta^*_0}\alpha + (1 - \alpha)\overline{\boldsymbol\theta}^*_1$, where the scaling $\alpha$ is determined by the line search method described here \url{https://inlabru-org.github.io/inlabru/articles/method.html}. This process is repeated until, for each parameter, the difference between two consecutive linearization points is less than $1\%$ of the marginal posterior standard deviation. The value $1\%$ is the default value used by the R-package \inlabru and can be changed by the user. Regarding $\boldsymbol\theta^*_0$ provided by the user, we suggest setting the parameters to a value which do not lead to extreme cases. In our experience, using $\boldsymbol\theta^*_0$ such that all the parameters are equal to 1 is a safe choice. Another option may be to set it equal to the maximum likelihood estimators. We recommend avoiding cases where parameters are equal, or very close, to zero (e.g. $<10^{-10}$), as well as far from it (e.g. $> 1000$), which may prevent the algorithm from converging.

The proposed method is implemented in \inlabru combining three Poisson models on different datasets. The reference to a Poisson model is merely artificial and used for computational purposes, it does not have any specific meaning. Specifically, we leverage the internal log-likelihood used for Poisson models by INLA (and \inlabru) to obtain the approximate Hawkes process log-likelihood. This is the only reason why we chose to implement our Hawkes process approximation using different Poisson models.

More formally, INLA has the special feature of allowing the user to work with Poisson counts models with exposures equal to zero (which should be improper). A generic Poisson model for counts $c_i, i = 1,...,n$ observed at locations $\mathbf x_i, i = 1,...,n$ with exposure $E_1,...,E_n$ with log-intensity $\log\lambda_P(\mathbf x) = f(\mathbf x, \boldsymbol \theta)$, in \inlabru has log-likelihood given by:

\begin{equation}
    \mathcal L_P(\boldsymbol \theta) \propto -\sum_{i = 1}^n \exp\{\overline{f}(\mathbf x_i, \boldsymbol \theta, \boldsymbol \theta^*)\}*E_i + \sum_{i=1}^n \overline{f}(\mathbf x_i, \boldsymbol \theta, \boldsymbol \theta^*)*c_i
    \label{eq:23_poissonloglik}
\end{equation}

Each Hawkes process log-likelihood component is approximated using one surrogate Poisson model with log-likelihood given by Equation \ref{eq:23_poissonloglik} and appropriate choice of counts and exposures data. Table \ref{tab:2_loglikapprox} reports the approximation for each log-likelihood component with details on the surrogate Poisson model used to represent it. For example, the first part (integrated background rate) is represented by a Poisson model with log-intensity $\log\Lambda_0(\mathcal X)$, this will be automatically linearised by \inlabru. Given that, the integrated background rate is just a scalar and not a summation, and therefore we only need one observation to represent it assuming counts equal 0 and exposures equal 1. Table \ref{tab:2_loglikapprox} shows that to represent a Hawkes process model having observed $n$ events, we need $1 + \sum_h (B_h) + n$ events with $B_h$ number of bins in the approximation of the expected number of induced events by observation $h$.

Furthermore, Table \ref{tab:2_loglikapprox} lists the components that has to be provided by the user, namely the surrogate Poisson models log-intensities. More specifically, the user only needs to create the datasets with counts $c_i$, exposures $e_i$, and the information on the events $\mathbf x_i$ representing the different log-likelihood components; and, to provide the functions $\log\Lambda_0(\mathcal X), \log\Lambda_h(b_{i,h}),$ and, $\log\lambda(\mathbf x)$. The linearisation is automatically performed by \inlabru as well as the retrieving of the parameters' posterior distribution. Regarding the functions representing integrals, they do not need to be exact, a function performing numerical integration is also fine.

\begin{table}
  \centering
  \scalebox{0.8}{
  \begin{tabular}{llllll}
    \toprule
    Name     & Objective   & Approximation & Surrogate $\log\lambda_P$ & Number of data points & Counts and Exposures \\
    \midrule
    Part I & $\Lambda_0(\mathcal X)$ & $\exp\overline{\log\Lambda}_0(\mathcal X)$ & $\log\Lambda_0(\mathcal X)$ & 1 & $c_i = 0$, $e_i = 1$       \\ \\
    Part II & $\sum_{h=1}^n\sum_{i = 1}^{B_h} \Lambda_h(b_{i, h})$ & $\sum_{h=1}^n\sum_{i = 1}^{B_h} \exp\overline{\log\Lambda}_h(b_{i,h})$ & $\log\Lambda_h(b_{i,h})$ & $\sum_h B_h$ & $c_i = 0$, $e_i = 1$       \\ \\
    Part III & $\sum_{h=1}^n\log\lambda(\mathbf x_h)$ & $\sum_{h=1}^n \exp\overline{\log\lambda}(\mathbf x_h)$ & $\log\lambda(\mathbf x)$ & $n$ & $c_i = 1$, $e_i = 0$       \\ \\
    \bottomrule
  \end{tabular} }
  \caption{Hawkes process log-likelihood components approximation}
  \label{tab:2_loglikapprox}
\end{table}

We provide a step-by-step tutorial on how to implement the approximation method described above. The tutorial gives more details on which functions has to be provided by the user, how to construct the binning strategy, how to set different priors for the parameters, and how to pass everything to \inlabru to retrieve the posterior distribution of the parameters. The tutorial can be found at \url{https://github.com/Serra314/Hawkes_process_tutorials/tree/main/how_to_build_Hawkes}. 

\section{Real Data Example}
\label{sec:5_realdata}

\begin{figure}[ht]
\centering
\includegraphics[width=.8\linewidth]{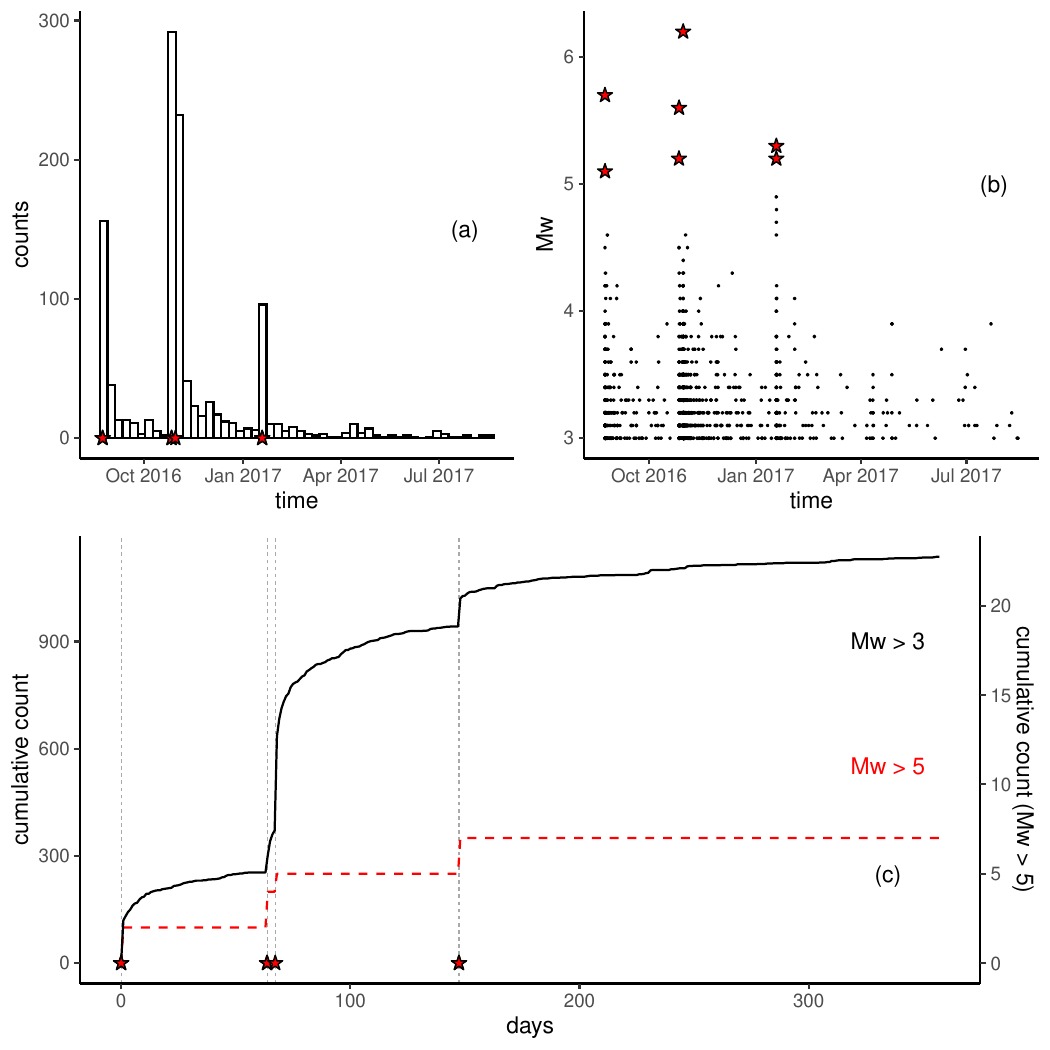}
\caption{Amatrice sequence comprising $1137$ events from 24/08/2016 to 15/08/2017, with longitude in $(42.45, 43.08)$ and latitude in $(12.93, 13.54)$. The first event in the catalogue is the magnitude 6.01 which started the sequence. Red stars indicate events with magnitude greater than 5. Panel (a): Histogram reporting the number of events per week; Panel (b): Scatter plot of time versus magnitude; (c) Cumulative number of events as function of the number of days from the first event in the sequence, for events with magnitude greater than 3 (solid black) and for events with magnitude greater than 5 (dashed red).}
\label{fig:1_amatrice}
\end{figure}

We provide a practical example of a temporal marked Hawkes process to illustrate the capabilities of our technique. We implement the temporal version of the Epidemic-Type-Aftershock-Sequence model (ETAS, \cite{ogata1988etas}), the most popular model to describe the evolution of seismicity in time, and we apply it to the 2016 Amatrice seismic sequence (\cite{marzocchi2017earthquake}). Specifically, we have considered $1137$ events with a magnitude greater or equal to $3$ from 24/08/2016 to 15/08/2017, with longitude in $(42.45, 43.08)$ and latitude in $(12.93, 13.54)$. The temporal evolution of the number of events is illustrated in Figure (\ref{fig:1_amatrice}). The data is taken from the Italian Seismological Instrumental and Parametric Database (ISIDe, \cite{iside}) downloaded from \url{https://doi.org/10.13127/ISIDE}.

The example consists of mainly two parts. In the first one, we compare the results of our implementation with the results obtained with the \bayesianETAS R-package (\cite{ross2021bayesian}), which provides an automatic MCMC implementation of the temporal ETAS model. The implementations are compared in terms of goodness-of-fit, expected number of events, and expected number of induced events. This is because we use different parameterizations preventing us from directly comparing the posterior of the parameters. We do this to show that our technique provides similar results to the MCMC implementations but in less time. This is relevant because we are working with an approximation method, while the MCMC implementation is exact, and the fact that both implementations provide similar results shows the accuracy of our approximation method.

In the second part of this example, we provide a retrospective daily forecasting experiment in which we compare daily forecasts of seismicity against observed seismicity in terms of number of events per day, for 120 days starting from 24/08/2016, just after the first large earthquake in the sequence. This is done using the \inlabru implementation only given the similarity of the results of the MCMC implementation. We use catalog-based forecasts (\cite{savran2020pseudoprospective}) for which the forecast for each day is composed of $10000$ simulated catalogs. Each simulated catalog is based on a different set of parameters extracted from the posterior distribution.

\subsection{ETAS model}
\label{sec:5.1_ETAS}

The ETAS model is the most used Hawkes process to model the evolution of seismicity over time and space (\cite{ogata1988etas, ogata2006space, ogata2011significant}). We are going to implement the first version of the model which is a temporal marked Hawkes process model with the event's magnitude as marking variable. The conditional intensity of the ETAS model is given by:

\begin{equation}
    \lambda_{\text{E}}(t,m | \mathcal H_t) = \left(\mu + K \sum_{h:t_h < t} \exp\{\alpha(m_h - M_0)\} (t - t_h + c) ^{-p} \right) \pi(m)
    \label{eq:24_ETAS}
\end{equation}

Where, $M_0 \in \mathbb R$ is the minimum recorded magnitude, and $\pi(m)$ is the magnitude distribution which is estimated independently from the Hawkes process parameters and assumed to follow a form of Gutemberg-Richter (GR) law (\cite{gutenberg1956magnitude}). The temporal evolution of the number of points is regulated by 5 parameters $\mu, K, \alpha, c, \geq 0$ and $p \geq 1$. The parameters $\mu, K,$ and $\alpha$ are productivity parameters regulating: the number of background events ($\mu$), the number of induced events or aftershocks ($K$), and how the aftershock productivity scales with magnitude ($\alpha$, the higher the magnitude the more events are generated). The parameters $c$ and $p$ are the parameters of the Omori's law (\cite{omori1894after}) and regulate the temporal decay of the aftershock activity. The quantity $M_0$ is a cut-off magnitude such that $m_h \geq M_0,\, \forall h$. 

The \bayesianETAS package implements the ETAS model with a normalized time triggering function to integrate to 1 over $(0, \infty)$. The conditional intensity is given by:

\begin{equation}
    \lambda_{\text{bE}}(t,m | \mathcal H_t) = \left(\mu + K \sum_{h:t_h < t} \exp\{\alpha(m_h - M_0)\} c^{p-1}(p-1)(t - t_h + c) ^{-p} \right) \pi(m)
    \label{eq:25_bEcondlambda}
\end{equation}

With our technique, it is best to work with a different parametrization than the one used in the \bayesianETAS package. Specifically, we choose the following conditional intensity

\begin{equation}
    \lambda_{\text{bru}}(t,m | \mathcal H_t) = \left(\mu_b + K_b \sum_{h:t_h < t} \exp\{\alpha_b(m_h - M_0)\} \left( \frac{t - t_h}{c_b} + 1\right)^{-p_b} \right) \pi(m)
    \label{eq:26_brucondlambda}
\end{equation}

The parameters of the \inlabru implementation have the same constraints, and the same interpretation, as in the \bayesianETAS implementation. The two implementations are equivalent considering  

\begin{equation}
    K_b = \frac{K(p-1)}{c}, \quad c_b = c, \quad p_b = p
     \label{eq:27_Kb}
\end{equation}

However, we are not going to use the above constraint in the example. The only constraints that we impose are $\mu, K, \alpha, c \geq 0$ and $p > 1$.

\subsection{Priors}
\label{sec:5.2_realpriors}

Priors are an essential part of the Bayesian approach. The \bayesianETAS package has fixed priors that cannot be changed. Specifically, they consider,

\begin{align}
    \mu & \sim \text{Gamma}(0.1,0.1) \nonumber\\
    K, \alpha, c & \sim \text{Unif}(0,10)
    \label{eq:28_priors}
    \\
    p & \sim \text{Unif}(1,10) \nonumber
\end{align}

This set of priors induces a prior on the parameter $K_b$, using Equation (\ref{eq:27_Kb}), with very light tails, highlighting how informative uniform priors may be (\cite{zhu2004counter}). We use the same set of priors except for $K_b$ for which we choose a log-normal distribution matching the $1$ and $99\%$ quantiles of the empirical distribution of $K_b$ obtained simulating 1000000 independent samples of $K, c, p$ from the priors in Equation (\ref{eq:28_priors}). We chose a log-normal distribution with mean and standard deviation of the logarithm equal to $-1$ and $2.03$. Table \ref{tab:3_priorKb} reports summary statistics of the \bayesianETAS prior for $K_b$ and the log-normal prior we chose to replicate it. The full set of priors used to replicate the \bayesianETAS priors are 

\begin{align}
    \mu_b & \sim \text{Gamma}(0.1,0.1) \nonumber\\
    K_b & \sim \text{LogN}(-1, 2.03) \nonumber \\ 
    \alpha_b, c_b & \sim \text{Unif}(0,10)
    \label{eq:29_inlabru_rep_priors}
    \\
    p_b & \sim \text{Unif}(1,10) \nonumber
\end{align}

We use this replicate set of priors to minimize the differences between the implementations which do not depend on the methodology used to find the posterior distribution of the parameters. We refer to this case as \inlabru replicate case.

\begin{table}[ht]
    \centering
  \begin{tabular}{llllllll}
    \toprule
    Implementation & Mean & St.Dev & 0.01q & 0.25q & Median & 0.75q & 0.99q \\
    \midrule
    \bayesianETAS & 11.854 & 3583.873 & 0.004 & 0.111 & 0.262 & 0.758 & 41.914 \\ 
    \inlabru & 2.887 & 22.482 & 0.003 & 0.094 & 0.368 & 1.447 & 41.367 \\ 
    \bottomrule \\
  \end{tabular}  \caption{Prior distribution summary statistics of parameters $K_b$ in the \bayesianETAS and \inlabru implementation. The distribution in the \bayesianETAS case is obtained sampling independently 1000000 times from $K, c \sim \text{Unif}(0,10)$, $p \sim \text{Unif}(1,10)$, and setting $K_b = K(p-1)/c$. The distribution in the \inlabru case is a log-normal distribution with mean and standard deviation of the logarithm equal to $-1$ and $2.03$ in order to match the extreme quantiles of the \bayesianETAS case.}
    \label{tab:3_priorKb}
\end{table}

We also consider a different set of priors that better reflects the scale of each parameter. For example, for the \inlabru implementation the parameters, $\mu$ and $c$ are on a very different scale than $K, \alpha,$ and  $p$. To reflect this piece of information through the prior, we use gamma priors for all parameters with different parameters reflecting the different scales. This information is usually available from previous studies of the same model. We use  

\begin{align}
    \mu_b & \sim \text{Gamma}(0.1, 1) \nonumber\\
    K_b & \sim \text{Gamma}(1, 0.5) \nonumber\\
    \alpha_b & \sim \text{Gamma}(1, 0.5) 
    \label{eq:31_lnpriors} \\
    c_b & \sim \text{Gamma}(0.1, 1) \nonumber\\
    p_b - 1 & \sim \text{Gamma}(0.1, 0.5) \nonumber 
\end{align}

Table (\ref{tab:4_priorcomp}) reports a comparison between summary statistics of \bayesianETAS priors and the gamma priors.

\begin{table}[]
    \centering
  \begin{tabular}{lllllllll}
    \toprule
    Name & Mean & St.Dev & 0.01q & 0.25q & Median & 0.75q & 0.99q & Implementation \\
    \midrule
    $\mu$ & 1 & 3.162 & 0.000 & 0.000 & 0.006 & 0.353 & 15.884 & \bayesianETAS \,\\ 
    $\mu$ & 0.1 & 0.316 & 0.000 & 0.000 & 0.001 & 0.035 & 1.588 &  \inlabru - Gamma \\ 
    $K_b$ & 11.854 & 3583.873 & 0.004 & 0.111 & 0.262 & 0.758 & 41.914 & \bayesianETAS\,\\ 
    $K_b$ & 2 & 2 & 0.020 & 0.575 & 1.386 & 2.773 & 9.210 & \inlabru - Gamma \\ 
    $\alpha$ & 5 & 2.88 & 0.1 & 2.5 & 5 & 7.5 & 9.9 & \bayesianETAS\,\\ 
    $\alpha$ & 2 & 2 & 0.020 & 0.575 & 1.386 & 2.773 & 9.210 & \inlabru - Gamma \\ 
    $c$ & 5 & 2.888 & 0.1 & 2.5 & 5 & 7.5 & 9.9 & \bayesianETAS\,\\ 
    $c$ & 0.1 & 0.316 & 0.000 & 0.000 & 0.001 & 0.035 & 1.588 &  \inlabru - Gamma \\ 
    $p$ & 5.5 & 2.598 & 1.09 & 3.25 & 5.5 & 7.75 & 9.91 & \bayesianETAS\,\\ 
    $p$ & 1.2 & 0.632 & 1.000 & 1.000 & 1.001 & 1.071 & 4.177 &  \inlabru - Gamma\\
    \bottomrule \\
  \end{tabular}  \caption{Prior distribution summary statistics of ETAS parameters for the \bayesianETAS implementation and the \inlabru gamma case which considers $\mu, c \sim \text{Gamma}(0.1, 1)$, $K, \alpha \sim\text{Gamma}(1, 0.5)$, and $p - 1 \sim \text{Gamma}(0.1, 0.5)$.}
  \label{tab:4_priorcomp}
\end{table}

In the remainder of the article, we refer to the \inlabru implementation considering the priors in Equation \ref{eq:29_inlabru_rep_priors} as \inlabru replicate and to the \inlabru implementation with the priors in Equation \ref{eq:31_lnpriors} as \inlabru gamma. Appendix \ref{sec:App.A_post} compares the prior and the posterior distributions for each model and shows the robustness of \inlabru's results under change of priors. Furthermore, Appendix \ref{sec:App.C_priorvar} provides a more complete prior sensitivity analysis. In there, we consider all the parameters as having the same log-normal prior, with the logarithmic mean equal 0 and different values of the logarithmic standard deviation.  

\subsection{Copula transformation}

The INLA method is designed for Latent Gaussian models and, therefore, all the parameters should have a Normal distribution. This is not the case for the ETAS parameters and the priors illustrated in the previous section. In order to overcome this problem we are going to use a copula transformation. Using this method allows us to represent internally the parameter as free-constraints and normally distributed. The constraints are implemented through the transformation itself. 

More formally, we use a transformation method based on the probability integral transform. The probability integral transform can be stated as follows: 

\begin{theorem}
Given a continuous random variable $X$ with cumulative distribution function (CDF) $F_X(\cdot)$, then the variable 

$$
Y = F_X(X)
$$

has a Uniform distribution in (0,1). 
\end{theorem}
 
The theorem implies also that given $Y \sim \text{Unif}(0,1)$ then, $X = F_X^{-1}(Y)$.
 
We apply this theorem by considering each parameter as having a standard normal distribution and then, transforming it to have the target distribution. More formally, assume $\theta$ has a starting distribution with CDF $F_\theta(\cdot)$, and that we want to transform it in $\eta(\theta)$ having a target CDF $F_Y(\cdot)$. Applying the transformation 

\begin{equation}
    \eta(\theta) = F_Y^{-1}\left(F_\theta(\theta)\right)
    \label{eq:30_linkf}
\end{equation}

the quantity $\eta(\theta)$ is distributed according to $F_Y$.

This allows us to consider a set of internal free-constraint parameters $\theta_\mu, \theta_K, \theta_\alpha, \theta_c, \theta_p$, representing (respectively) $\mu, K, \alpha, c, p$, with a standard normal prior distribution and then transforming them to have the desired prior distribution. We can incorporate the constraint on the parameter values using appropriate prior distributions. For example, using any distribution with positive support ensures that the transformed parameter is greater or equal to zero.

\subsection{Goodness-of-fit}
\label{sec:5.3_comp}

We compare the \inlabru and the \bayesianETAS implementation in terms of goodness-of-fit, this is due to the use of different parametrizations. Indeed, different parametrizations and different priors make a direct comparison of the posterior of the parameters elusive, because it is hard to determine if the differences in the posterior distributions come from the different parameterizations, the different priors, or the different methodologies. With this section, we want to convince the reader that our approximation provides results similar in terms of goodness-of-fit to MCMC implementations but in less time. This is relevant considering that MCMC is an exact method, with the ability to sample from the true marginal posteriors of the model, while our method is based on a series of approximations. Showing that the \inlabru implementation provides similar results shows the goodness of the approximation.

We compare the goodness-of-fit of the models using the Random Time Change Theorem (\cite{meyer1971demonstration}). This is a standard technique to measure the goodness-of-fit for Hawkes process models as described in \cite{laubelements}. Below we report the Random Time Change Theorem as stated in \cite{laubelements} (Theorem 9.1): 

\begin{theorem}
Say $\mathcal H = \{t_1,...,t_k\}$ is a realisation over time $[0,T]$ from a point process with conditional intensity $\lambda(t | \mathcal H)$. If $\lambda(t|\mathcal H)$ is positive over $[0,T]$ and $\Lambda(T) < \infty$ almost surely, then the transformed points $\{\Lambda(t_1),...,\Lambda(t_k)\}$ form a Poisson process with unit rate.
\end{theorem}

Where in our case,

\begin{equation}
    \Lambda(t_i | \mathcal H) = \int_{M_0}^{\infty} \int_{0}^{t_i} \lambda(t, m |\mathcal H) dt dm
    \label{eq:32_trasnf}
\end{equation}

In other words, if we calculate the sequence of values $\Lambda(t_1),...,\Lambda(t_n)$, for observed $t_1,...,t_n$, using the respective expressions of $\Lambda(t_i)$ for the \bayesianETAS and \inlabru implementation, we have to obtain a sequence of points uniformly distributed over the interval $[0, n]$, where $n$ is the number of observed points. For the MCMC method, we consider estimates based on 10000 posterior samples with a burn-in of 5000 samples. The \bayesianETAS package requires around 9 minutes to generate a total of 15000 posterior samples, while the \inlabru method only requires around 3 minutes to converge. Section \ref{sec:6_sim} shows how these times scales increasing the number of observations, while Appendix \ref{sec:App.B_binning} illustrates the variation of the \inlabru computational time for different binning strategies.

\begin{figure}[ht]
\centering
\includegraphics[width=.8\linewidth]{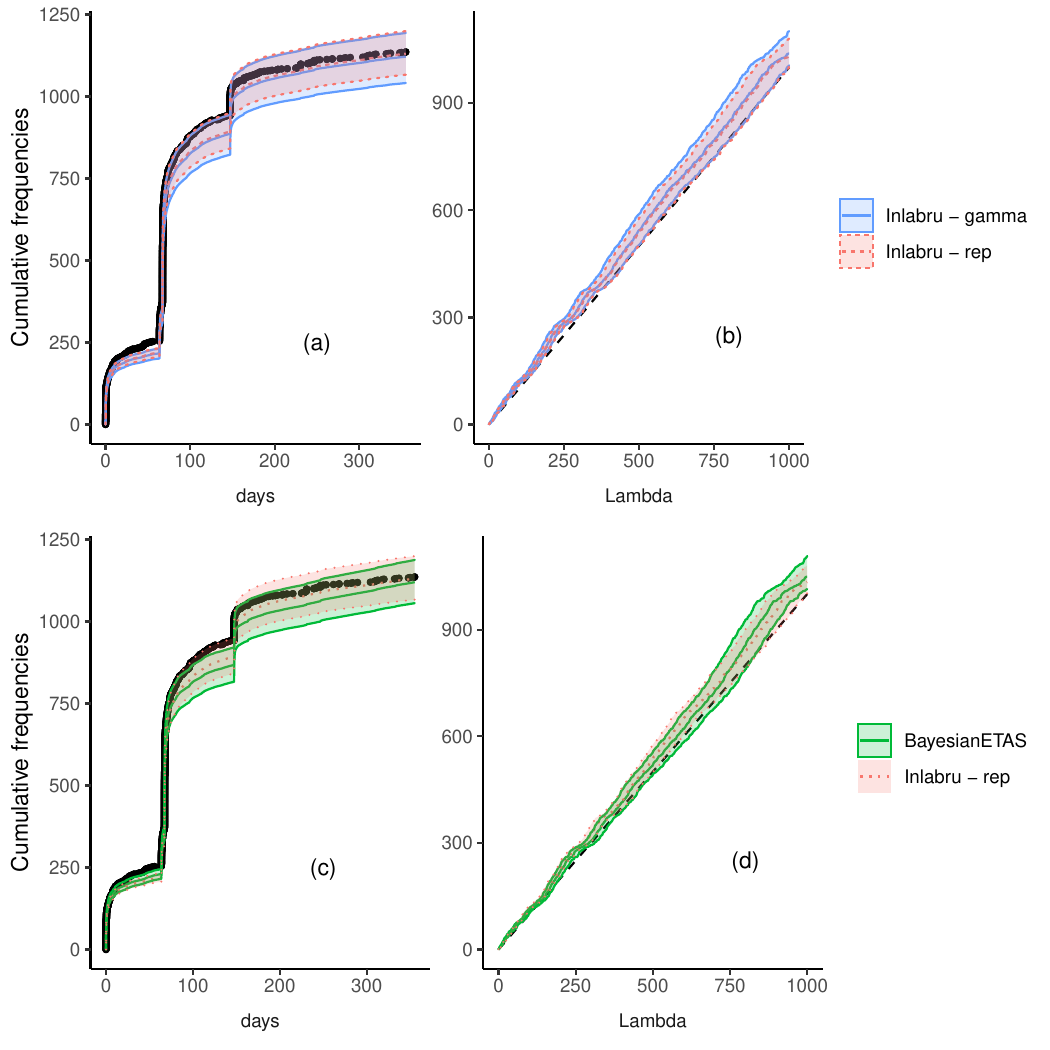}   
\caption{Application of the Random Time Change Theorem. Top row (a-b): Compares the \inlabru replicate and gamma (solid blue) implementations. Panel a-c: Observed cumulative number of events as a function of time (black dots) with the prediction provided by the model; Bottom row (c-d): Compares the \bayesianETAS (solid green) and \inlabru replicate (dotted red) implementations; Panel a-c : Cumulative number of events as a function of time. Panel b-c : Cumulative number of events as a function of $\Lambda(t_h)$, the black dashed line represents the uniform case. The shaded region represents the $95\%$ predictive interval for each quantity obtained by sampling 10000 times the posterior of the parameters.}
\label{fig:2_rtctheo}
\end{figure}

Figure \ref{fig:2_rtctheo}a-c compares the sequences $\Lambda_\text{bE}(t_1), ... \Lambda_\text{bE}(t_n)$, and $\Lambda_\text{bru}(t_1), ... \Lambda_\text{bru}(t_n)$ with observed cumulative counts $N(t_1),...,N(t_n)$.  Figure \ref{fig:2_rtctheo}b-d shows the cumulative counts as a function of $\Lambda(t_h)$ and should look like a straight line if the values are uniformly distributed as expected by the theorem. For both plots, we report $95\%$ posterior intervals for the quantity of interest based on 10000 samples from the posterior of the parameters.

There are small differences between the two \inlabru implementations, which was expected from the similarity of the posterior distributions provided by the model and reported in Figure \ref{fig:9_post_brucomp_log10}. The differences in the results are greater if we compare the \bayesianETAS and the \inlabru implementations. In fact, the \inlabru implementation estimates a slightly lower background rate and a greater capability of each event of generating aftershocks, which allows the prediction to match the observations in the last part of the sequence. In fact, in Figure \ref{fig:2_rtctheo} (d) the dashed line representing the theoretical uniform distribution is outside the \bayesianETAS boundaries while it is inside the \inlabru ones. Apart from these small differences, the three implementations provide consistent results. 

The main difference between the \bayesianETAS and \inlabru implementations is the computational time. The \bayesianETAS R-package requires around $4, 6, 9$ minutes to generate, respectively, $1000, 5000, 10000$ posterior samples considering $5000$ burn-in samples. Our \inlabru implementations require around $3$ minutes to converge for different binning strategy. The minimum convergence time is $2.93$ minutes obtained, while the maximum is $3.7$. Table \ref{tab:5_bins} reports the computational time and iterations needed for convergence for different binning strategy parameters.

\subsection{Expected number of events and branching ratio}
\label{sec:5.4_NBR}

We also compare the \inlabru and \bayesianETAS implementations in terms of the expected number of events and branching ratio. This is done because these two quantities are usually relevant in applications. Given a Hawkes process model with conditional intensity $\lambda(t | \mathcal H_t)$, the expected number of events in a time interval $(T_1, T_2)$, $0 \leq T_1 < T_2 < \infty$ given the history of the process is given by the integral of the conditional intensity 

\begin{equation}
    \Lambda(T_1, T_2) = \int_{T_1}^{T_2} \lambda(t | \mathcal H_t) dt
\end{equation}

The number of points has a Poisson distribution with rate $\Lambda(T_1, T_2)$. 

\begin{figure}[ht]
\centering
\includegraphics[width=.9\linewidth, height = .7\linewidth]{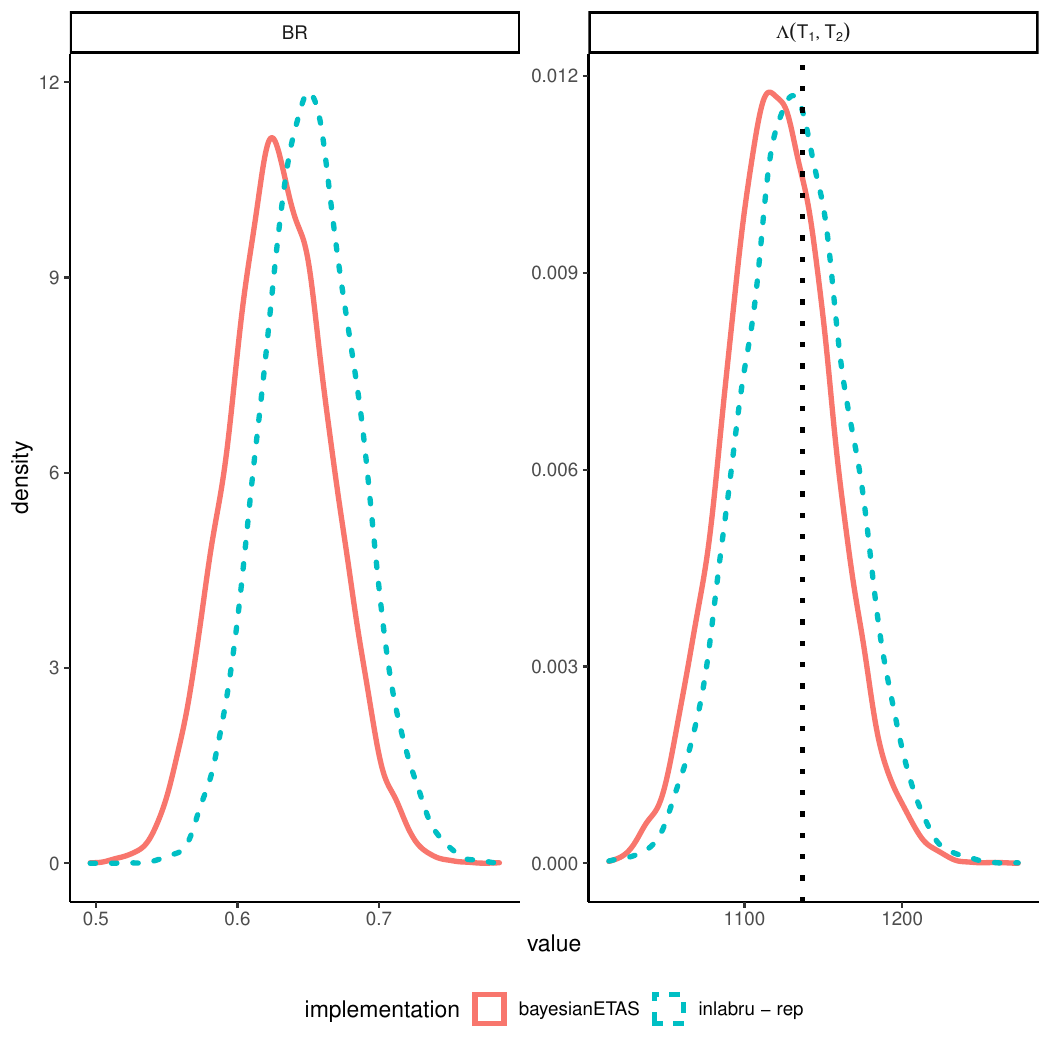}  
\caption{Right Panel: Expected number of events $\Lambda(T_1, T_2)$ posterior distribution comparison for the \inlabru replicate implementation (blue dashed) and the \bayesianETAS implementation (red solid). The vertical dotted line represents the observed number of points. Left Panel: Branching ratio $\text{BR}$ posterior distribution comparison for the \inlabru replicate implementation (blue dashed) and the \bayesianETAS implementation (red solid).}
\label{fig:3_comp.NBR}
\end{figure}

Figure \ref{fig:3_comp.NBR} (right) shows the posterior distributions of $\Lambda(T_1, T_2)$ for the \inlabru and \bayesianETAS implementations. We show only the \inlabru replicate case given that the \inlabru gamma case provides the same results. For the two implementations, the posterior distribution of $\Lambda(T_1, T_2)$ is estimated by calculating the analytical expression of $\Lambda(T_1, T_2)$ for the two approaches using $10000$ samples from the posterior distribution of the parameters. The approaches provide coherent results between each other, although the mode of the posterior distribution of $\Lambda(T_1, T_2)$ is closer to the observed number of points (vertical dashed line) in the \inlabru case.

Another important quantity in analyzing Hawkes process models is the branching ratio $\text{BR}$. The branching ratio is the expected total number of events induced by another event. The branching ratio can be calculated as the integral of the excitation (triggering) function for time differences going from 0 to $\infty$. In the ETAS case, we have an excitation function that depends also on the magnitude, namely $g: (0, \infty) \times (M_0, \infty) : \rightarrow (0, \infty)$ such that

\begin{equation}
    g(t - t_h, m_h) = g_t(t - t_h, m_h) \pi(m_h)
\end{equation}

where $\pi(m_h)$ is the magnitude distribution. 

In this case, the branching ratio is given by

\begin{equation}
    \text{BR} = \int_{M_0}^{\infty} \left(\int_{0}^{\infty} g_t(s, m)ds \right)\pi(m)dm  
\end{equation}

Therefore, the branching ratio can be seen as the expected value under the magnitude distribution of the expected number of events induced by another. Assuming to have a point in 0, then the number of points induced by that event has a Poisson distribution with rate $\sum_{i = 1}^{\infty} \text{BR}^i$. As explained by \cite{laubelements} in Section 3 the branching ratio should be between 0 and 1 for the process to be stationary and for asymptotic results to be valid (\cite{hawkes1971spectra}). We did not set any constraint to ensure this property in the present implementation.

To calculate the branching ratio for a given set of parameters, we calculate analytically the inner integral $10000$ times, using samples from the magnitude distribution and we take the mean. This is repeated for $10000$ times, using as ETAS parameters samples from the parameters' posterior distribution. In this way, we obtain $10000$ samples from the posterior distribution of the branching ratio which can be used to approximate the posterior distribution empirically. Figure \ref{fig:3_comp.NBR} (left) compares the posterior distributions of the branching ratio for the \inlabru and \bayesianETAS implementations. Both posterior distributions only assign a positive probability to value between $0$ and $1$. The one obtained with \inlabru has a slightly smaller posterior variance and a larger mode. This is due to the smaller background rate estimated by the \inlabru implementation which in turns imply a higher number of induced events.

\subsection{Retrospective Forecasting Experiment}
\label{sec:5.5_retrofore}

We perform a retrospective daily forecasting experiment using the same data used to fit the data on the Amatrice seismic sequence. Specifically, for each forecasting period defined by $(t_j, t_{j+1})$, we simulate $10000$ synthetic catalogs assuming known all the events happened strictly before the forecasting period, namely $\mathcal H_{t_j}$. If, in the forecasting period $(t_j,t_{j+1})$ there is an earthquake with magnitude greater than $5.5$ with recorded time $t_m : t_j < t_m < t_{j+1}$, then, we consider the forecast for the period $t_j, t_m$ and we start a new daily forecast from $t_m + dt$, for $dt > 0$ (we use $dt = 10^{-6}$ days). This is done to resemble a true forecasting experiment, like the ones performed by the Collaboratory for the Study of Earthquake Predictability (CSEP, \cite{savran2020pseudoprospective} and reference therein), in which the forecasts are updated in presence of large earthquakes.

\begin{figure}[ht]
\centering
\includegraphics[width=.8\linewidth]{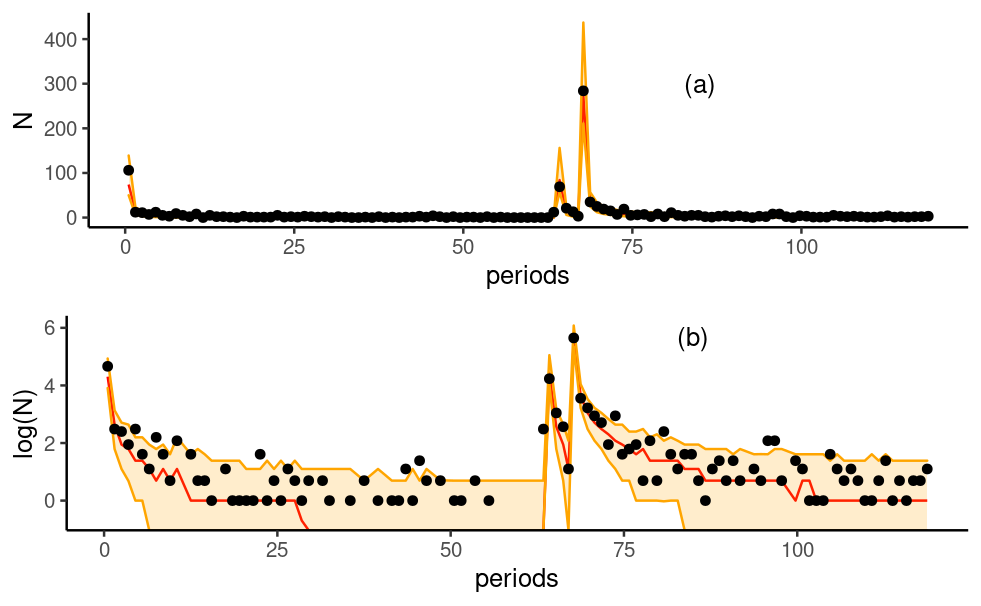}   
\caption{Retrospective forecasting experiment results. Black dots represent the observed number of events per forecasting period; the red solid line represents the median of the number of events of the synthetic catalogs per forecasting period; the shaded region represents the $95\%$ forecasting intervals for the number of events of the synthetic catalogs per forecasting period. Panel (a) shows the number of events in the natural scale. Panel (b) shows the logarithm of the number of events, periods with zero events have been omitted.}
\label{fig:4_fore_exp}
\end{figure}

The results of the retrospective experiment are shown in Figure \ref{fig:4_fore_exp}. The shaded region represents the $95\%$ forecasting interval of the number of events for each period. The extremes of each interval are the $2.5\%$ and the $97.5\%$ quantiles of the number of events of the synthetic catalogs composing the forecast for each day. Almost all observed numbers of events are comprises in the forecasting interval, particularly, all the periods with more than 50 events are correctly predicted. This is particularly relevant for applications on hazard/risk analyses in which the focus are on the periods just after large earthquakes where damages occur. 

\section{Simulation Experiment}
\label{sec:6_sim}

We performed a simulation example to compare the robustness of the \inlabru and \bayesianETAS approach if applied to different catalogs coming from the same model, and to give an idea on how the computational time scales increasing the amount of data. As data generating model, we use the \inlabru replicate implementation presented in Section \ref{sec:5.2_realpriors}. We generate 10000 synthetic catalogs for the period going from 24/08/2016 to 15/08/2017 (same period used for the Amatrice sequence) using as parameters the posterior median. In simulating the catalogs, we assume as known the 3 events with the greatest magnitude in the Amatrice catalogue recorded, respectively, on the 24/08/2016, 26/10/2016, and 30/10/2016, with magnitudes 5.7, 5.6, and 6.2. This is done to have a high probability of having, at least, $800$ events per catalog.  From the set of synthetic catalogs we select $5$ catalogs corresponding to $900, 1500, 2000, 2500, 3500$ number of events. We use these catalogs to fit $5$ different models with the \inlabru and \bayesianETAS implementations. For the \inlabru implementation, we use the same priors and starting points as in the \inlabru replicate case and binning strategy parameters given in Appendix \ref{sec:App.B_binning}. For the \bayesianETAS implementation, we consider $5000$ posterior samples with $5000$ burn-in samples.

\begin{table}[]
    \centering
  \begin{tabular}{llll}
    \toprule
    N events & \bayesianETAS & \inlabru & time ratio \\
    \midrule
    900 & 3.90 (mins) & 2.96 (mins) & 1.31 \\ 
    1500 & 9.75 (mins) & 1.56 (mins) & 6.21 \\ 
    2002 & 16.80 (mins) & 2.69 (mins) & 6.24 \\ 
    2500 & 30.73 (mins) & 2.75 (mins) & 11.15 \\ 
    3500 & 56.09 (mins) & 5.22 (mins) & 10.72 \\ 
    \bottomrule \\
  \end{tabular}  \caption{Comparison of computational times for the \bayesianETAS and \inlabru implementations in minutes. Last column report the ratio between the number of minutes needed by \bayesianETAS and \inlabru.}
  \label{tab:5_comptime}
\end{table}

\begin{figure}[ht]
\centering
\includegraphics[width=.8\linewidth]{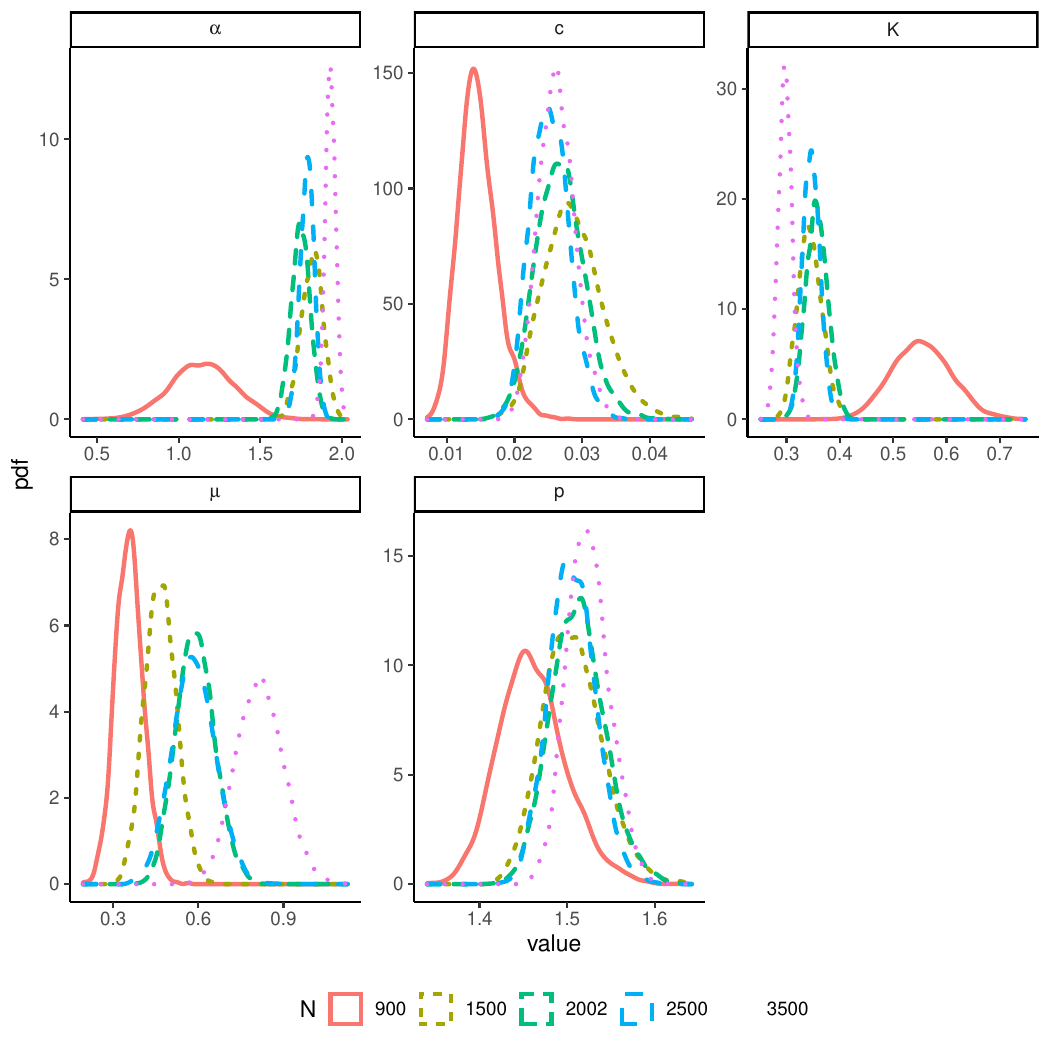}   
\caption{ETAS parameters' posterior distribution using the \bayesianETAS R-package on $5$ synthetic earthquake catalogs. The color and the linetype represents the number of events in each synthetic catalog. The synthetic catalogs are simulated using as parameters the median of the posterior distribution of the \inlabru replicate implementation obtained on the Amatrice seismic sequence.}
\label{fig:5_sim.mcmc}
\end{figure}

\begin{figure}[ht]
\centering
\includegraphics[width=.8\linewidth]{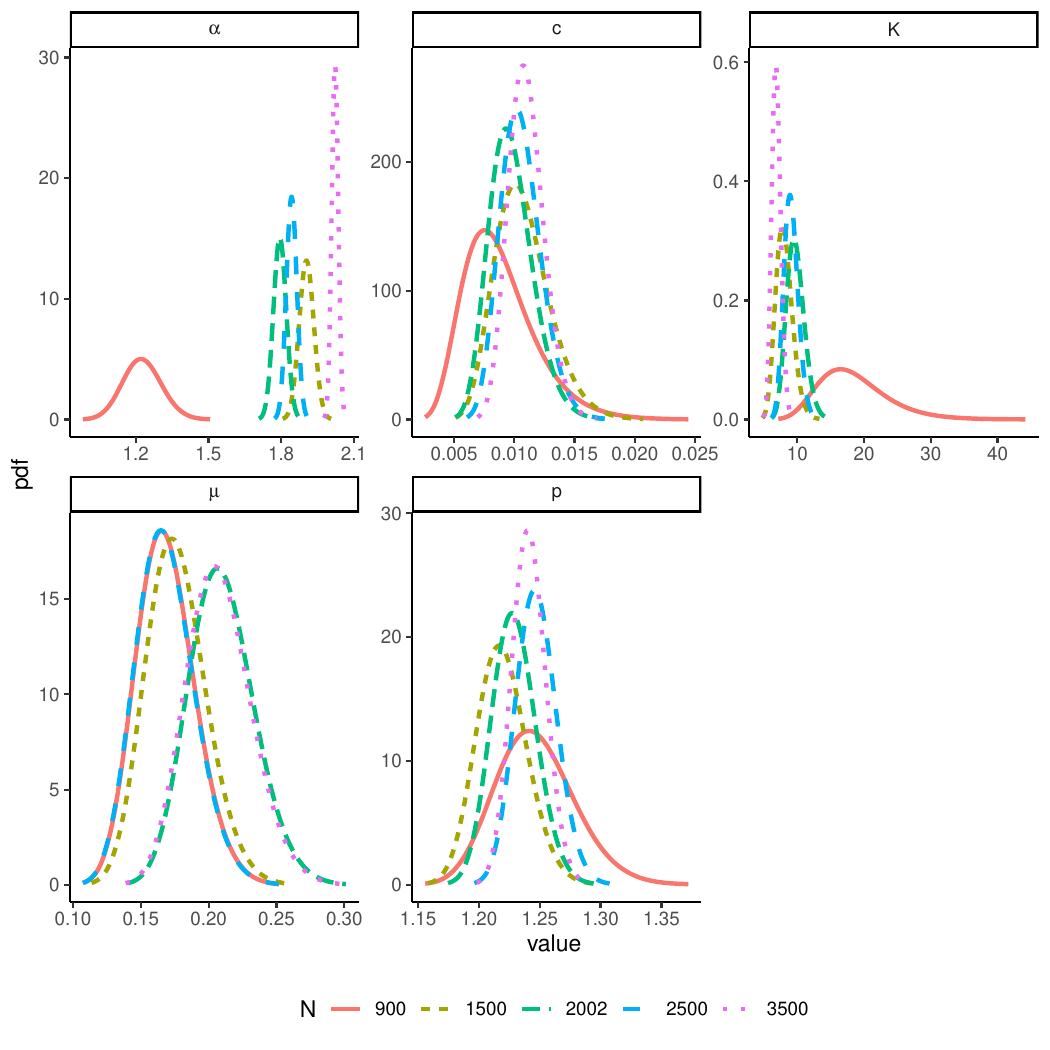}   
\caption{ETAS parameters' posterior distribution using the \inlabru R-package on three synthetic earthquake catalogs. The color and the linetype represents the number of events in each synthetic catalog. The synthetic catalogs are simulated using as parameters the median of the posterior distribution of the \inlabru replicate implementation obtained on the Amatrice seismic sequence.}
\label{fig:6_sim.bru}
\end{figure}

Table \ref{tab:5_comptime} shows how the computational time scales increasing the number of events in the data for the two implementations. The advantages of the \inlabru approach are clear, especially for catalogs with more than $2500$ events for which \inlabru is $10$ times faster than \bayesianETAS. Figure \ref{fig:5_sim.mcmc} (\bayesianETAS) and \ref{fig:6_sim.bru} (\inlabru) show the posterior of the parameters for the different simulated catalogs. The differences between the posteriors obtained by each approach on different catalogs are expected. For example, the case with $3500$ (as well as $900$) events can be considered an extreme case and, thus, the posterior distribution would be different from more \emph{common} catalogs. Indeed, the parameters $\mu, K, \alpha$, regulating the number of events, are the ones with more differences in the posteriors for different catalogs, while the parameters $p$ and $c$ regulating the temporal decay of the induced events are more similar. In this regard, the \inlabru implementation is more stable than the \bayesianETAS implementation providing posteriors distributions more similar between each other. This is particularly true for parameters $\mu, c,$ and $p$. In addition, the two implementations provide coherent results between each other, for example, analyzing parameter $\alpha$, for both approaches the parameter's posterior distribution moves to the right as we increase the amount of data, and the opposite happens for parameter $K$.  

The coherence of the results for the two implementations considered illustrates the reliability of our approximation, and, the gain in computational time shows the advantage of our approach. Furthermore, the gain in computational time would be even greater if more complex models are considered. For example, we foresee that the computational gain will increase considering a spatio-temporal model, or, alternatively, considering one of the parameters as temporally varying. This has not to be underrated, in fact, in seismology, many researchers are discouraged to update their models (in an online fashion) or using large catalogs ($>100000$ events) by the price to pay in terms of computational time.

\section{Discussion and conclusions}
\label{sec:7_discuss}

In this paper, we presented a technique to implement Bayesian Hawkes process models based on the INLA algorithm and carried out with the R-package \inlabru. The proposed technique is new and differs substantially from other Hawkes process implementations. Specifically, we rely on a new Hawkes process log-likelihood approximation technique which allows us to apply the INLA method to Hawkes process models. Our technique provides similar results, in terms of goodness-of-fit, expected number of events, and branching ratio, as an MCMC technique (\cite{ross2021bayesian}) implemented through the \bayesianETAS package, but requiring less time. Regarding the time, the \bayesianETAS approach requires around double the time required by our technique for catalogs composed of circa 1000 events, and 10 times more for catalogs with more than 2500 events. We believe that in more complex cases (e.g spatio-temporal case, inclusion of covariates, parameters with structured variations) the gain in computational time provided by the \inlabru approach would be even larger. We have also shown that our technique provides reasonable results in a retrospective forecasting experiment, correctly predicting the number of events per day for most of the considered days. Furthermore, our algorithm is deterministic ensuring the same numerical results if the analysis is repeated on different machines with the same specifics. Moreover, the user does not have to program explicitly the algorithm itself, they only have to provide the functions to be approximated, and the approximation is performed automatically by the \inlabru R-package. Also, we do not rely on any declustering algorithm assigning the observations to the background rate or the triggered part of the intensity. 

An important difference from other algorithms for Hawkes process models is that we offer a general extendible framework to perform Bayesian analyses of Hawkes process parameters. Indeed, INLA was designed for models comprising covariates and random effects, and to compare them. This allows us to bring the advantages of the Latent Gaussian model world into the Hawkes process world. For example, we can consider the parameters as linear functions of available covariates. Another extension consists of considering the parameters as structured random effects: a parameter assumed to be a Gaussian Markov Random Field (GMRF) varying over space, or time, or both. For example, considering a parameter as an SPDE effect (\cite{lindgren2011explicit}) we can have spatially (or temporally) varying parameters where the absolute value of the correlation between the parameter's values at different locations (times) is a decreasing function of the distance between locations (times). Given the correlation between the parameter's values and the correlation between different parameters, these models would be difficult to implement using an MCMC technique, which, in case, should be tailored to the specific problem. On the other hand, INLA was designed specifically to efficiently handle large GMRF and correlated parameters. Using our method, all the models undergo the same optimization routine making them homogeneous under these aspects. When comparing two models optimized with different routines, it is hard to distinguish if the differences come from the different models or the different algorithms. Using our technique, researchers may compare models incorporating different hypotheses being sure of no differences, at least, on the optimization part, and thus, any difference in performance comes from the model formulation itself. 

The limitations of our approach reside in the functional form of the triggering (or excitation) function and the binning strategy. Specifically, we want the triggering function so that the quantities to be approximated are as close as possible to linearity. In our experience, the unnormalized version works best. Also, care has to be taken on the numerical stability of the provided functions which may be eased by linearly approximating them for values of the argument above/below a certain threshold. The binning strategy to further decompose Part II of the log-likelihood is essential to reach convergence. In our experience, a number of bins greater than 3 per observation is required. Also, the width of the bins is essential, considering too large bins prevents the algorithm to converge as shown in Table \ref{tab:5_bins}. We suggest to regulates the width and number of bins based on the problem at hand. For example, a triggering function decaying slowly with time would need larger bins than a function with a faster decay. With the same rationale, the function decaying slower needs fewer bins to be accurately approximated than one decaying faster.

Future developments will regard the inclusion of covariates and random effects in the model. We think that providing researchers with the freedom of focusing on the hypotheses incorporated in the model, and not on the optimization routine, is essential, especially in applied contexts. To facilitate the use of our technique, we are working on a R-package to automatically fit a Hawkes process model, retrieve information on the parameters' posterior distribution, and produce forecasts. We are planning to start with a R-package focused on the ETAS model and extend it to include different Hawkes process models. Indeed, we have already provided these functions in a tutorial \footnote{The tutorial is available at \url{https://github.com/Serra314/Hawkes_process_tutorials/tree/main/how_to_use_Hawkes}}. Specifically, we provided the user with one-line functions to fit the ETAS model used in the real data example on user-specified datasets, retrieve the posterior distributions of the parameters and the number of points, and produce forecasts for a user-specified number of periods and period's length. We have also made publicly available another tutorial\footnote{The tutorial is available at \url{https://github.com/Serra314/Hawkes_process_tutorials/tree/main/how_to_build_Hawkes}} illustrating in detail how to build the functions used in the first tutorial. The second tutorial explains which functions have to be provided by the user, how to construct the binning strategy, and how to make them interact with \inlabru and provides details on the possible difficulties that may be encountered in each step. This can be used as a template to implement Hawkes process models different from the ETAS model. 

To conclude, we have shown that the \inlabru approach is a valuable alternative to MCMC techniques for Hawkes process models, it provides comparable results in terms of quality but in a fraction of the time needed by MCMC. This is particularly relevant in applied contexts, such as seismology, where researchers are discouraged to use Hawkes process models on large datasets ($> 100000$ observations) by long computational times. On the same line, models used to produce daily forecasts are not updated daily, for the same reasons. The \inlabru approach softens this burden and allows researchers to fit models on larger datasets in less time. Also, our approach can be extended to consider more complex models which would have needed an ad-hoc implementation if an MCMC technique had to be used. We believe that the \inlabru approach could make Hawkes models more accessible for a greater number of users, which would have the freedom to make inference on                          models incorporating different hypotheses without the burden of adapting the methodology.  

\section{Data availability}
\label{sec:8_data}

The code and the files needed to reproduce all the results of the paper can be found at \url{https://github.com/Serra314/Hawkes_process_tutorials/tree/main/code_for_paper}. The two tutorials on how to use and how to implement Hawkes process models with \inlabru can be found at \url{https://github.com/Serra314/Hawkes_process_tutorials}. 
For the real data example, we used the Italian Seismological Instrumental and Parametric Database (ISIDe, \cite{iside}) which can be downloaded from \url{https://doi.org/10.13127/ISIDE}.

\section{Acknowledgements}

This research was supported by the European Union H2020 program (No 821115, Real-time earthquake rIsk reduction for a reSilient Europe
“RISE”, http://www.rise-eu.org/home/). For the purpose of open access, the author has applied a Creative Commons Attribution (CC BY) licence to any Author
Accepted Manuscript version arising from this submission. All the code to
produce the present results is written in the R programming language. We have used the package \texttt{ggplot2} (\cite{wickham2016ggplot2}) for all the plots in this manuscript.

\printbibliography
\newpage
\appendix

\section{Appendix A: parameters posterior distribution}
\label{sec:App.A_post}

\begin{figure}[ht]
    \centering
    \includegraphics[width = 15cm, height = 12cm]{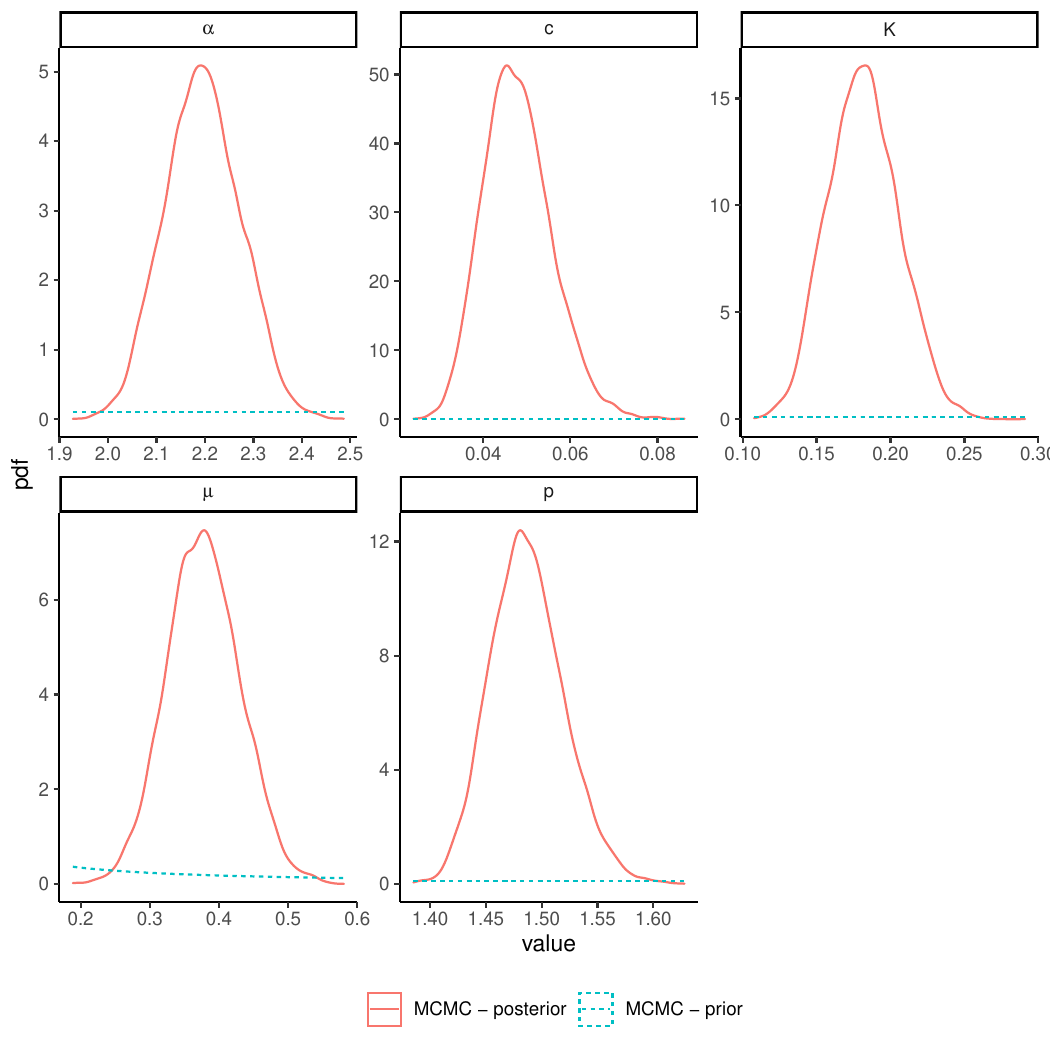}
    \caption{Posterior and prior distributions of ETAS parameter using the \bayesianETAS package considering 1000 posterior samples and 5000 burn-in samples. The results are based on the Amatrice seismic sequence.}
    \label{fig:7_post_mcmc}
\end{figure}

\begin{figure}[ht]
    \centering
    \includegraphics[width = 15cm, height = 12cm]{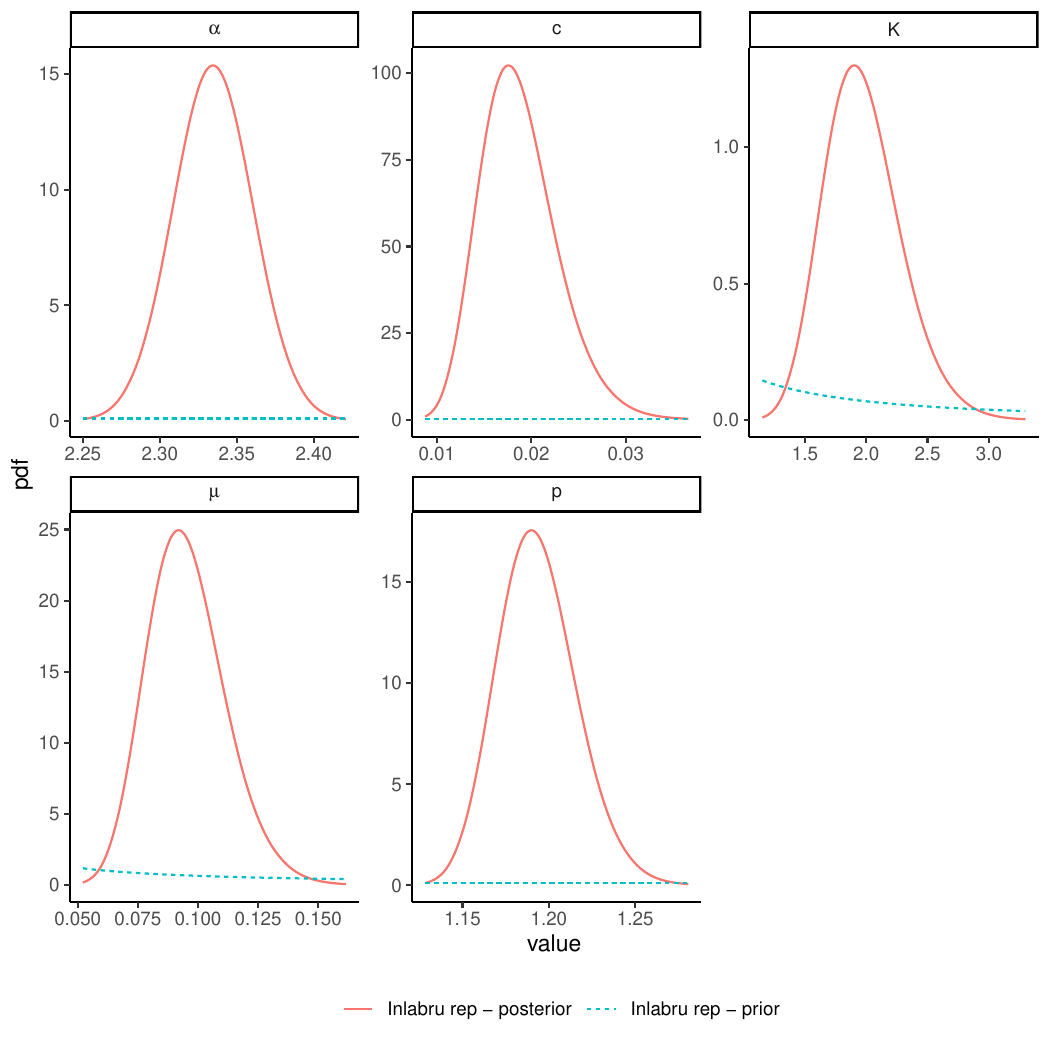}
    \caption{Posterior and prior distributions of ETAS parameter for \inlabru replcate case. The results are based on the Amatrice seismic sequence.}
    \label{fig:8_post_brucomp}
\end{figure}

\begin{figure}[ht]
    \centering
    \includegraphics[width = 15cm, height = 12cm]{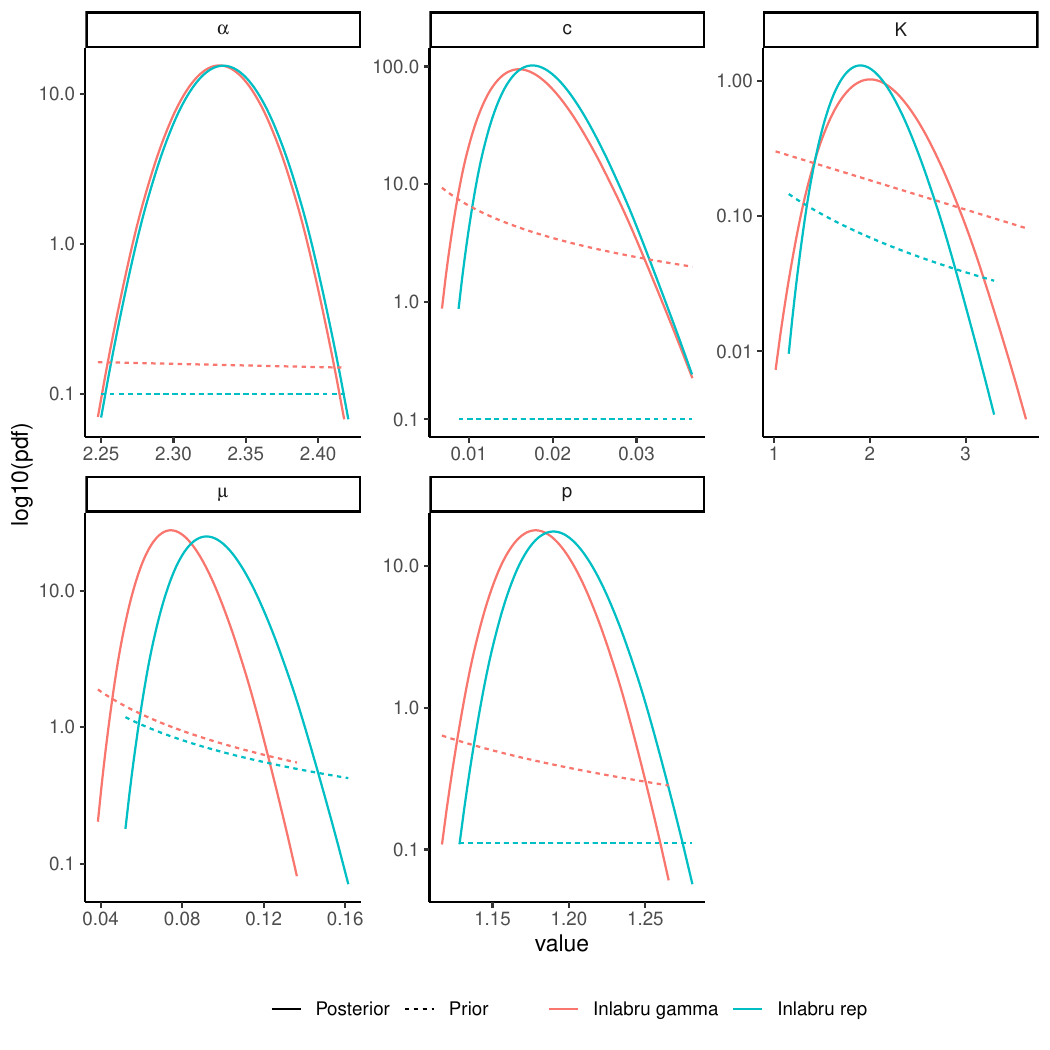}
    \caption{Posterior and prior distributions of ETAS parameter for the two \inlabru implementations considered, namely gamma and replcate. The value of the density is on a logarithmic (base 10) scale to highlight the differences in the prior.}
    \label{fig:9_post_brucomp_log10}
\end{figure}

Here, we show the marginal posterior distribution of the ETAS parameters calibrated on the Amatrice sequence comprising 1137 events from 24/08/2016 to 15/08/2017 with latitude in $(42.456, 43.084)$, and longitude in $(12.936, 13.523)$.  Below are reported the posterior distribution of the ETAS parameters for the implementations considered in the article. Figure \ref{fig:7_post_mcmc} shows the posterior distributions obtained using the MCMC implementation provided by the R-package \bayesianETAS considering 10000 posterior samples and 5000 burn-in samples. Figure \ref{fig:8_post_brucomp} shoes the posterior distribution of the ETAS parameters for the \inlabru replicate case, while Figure \ref{fig:9_post_brucomp_log10} compares the distribution of the \inlabru replicate and gamma implementations. For the latter, we chose to use a logarithmic scale for the comparison to highlight the differences in the prior.

\newpage
\section{Appendix B: Sensitivity to binning strategy}
\label{sec:App.B_binning}

 In our three factors decomposition of the point process log-likelihood, to approximate the second part (the expected number of triggered events Sec \ref{sec:4.2_part2}), we split the time domain into bins and we approximate the integral in each bin separately. In this paper, we use a different set of bins for each observed point. Specifically, for each arrival time $t_h$, the bins are defined by the sequence: 
 
 \begin{equation*}
     t_h, t_h + \Delta, t_h + \Delta(1 + \delta), t_h + \Delta(1 + \delta)^2, ...., t_h + \Delta(1 + \delta)^{n_h},  T_2
 \end{equation*}
 
 where $n_h$ is such that $t_h + \Delta(1 + \delta)^{n_h} < T_2$ or $n_h < n_{max}$. This binning strategy is defined by three parameters: $\Delta$ regulating the length of the first bin, $\delta$ regulating the increase in length of each subsequent bin, and $n_{max}$ which regulates the maximum number of bins per observed points ($n_{max} + 2$). 
 
In this section, we take the \inlabru replicate implementation and we try different parameters of the binning strategy. Specifically, we consider $\delta = 1,2,3,4,5,7$, $\Delta = 0.1, 0.2, 0.5$ and $n_{max} = 3, 10$. The binning strategy affects mostly the ability to converge and the computational time required to reach convergence. Table \ref{tab:5_bins} reports the number of iterations needed for convergence (n iter), the computational time (in minutes), and the convergence state for each combination of binning strategy parameters. We set a maximum number of iterations equal to 100 so that if the number of iterations for convergence is equal to 100 it means that the algorithm has not converged. We checked that the models are not able to converge looking at the posterior modes for each iteration of the algorithm, more detail on how to retrieve these quantities are reported in the tutorial on how to implement Hawkes process models with \inlabru. The fact that different binning strategies converge in a similar number of iterations highlights the robustness of our approach. The time needed for each iteration changes with different binning strategies.

Examining Table \ref{tab:5_bins}, models with $\delta = 7,10$ tend to not converge. This is due to the fact that these binning strategies induce too wide bins (especially close to the observations, where we need a finer partition) which in turn provide an approximation that is not accurate enough. Instead, strategies with $\delta = 2$ behave well and are the fastest to converge. In this paper, we use a binning strategy defined by $\delta = 2$, $\Delta = 0.1$ and $n_{max} = 3$ because it is the fastest to reach convergence.

\begin{table}[!htbp]
    \centering
  \begin{tabular}{llllll}
    \toprule
    $\delta$ & $n_{max}$ & $\Delta$ & n iter & time (mins) & converged \\
    \midrule
      2 &  3  & 0.2  &     63 & 2.93  &    TRUE \\
      2 &  10  & 0.2  &     63 & 2.98  &    TRUE \\
      2 &    10 &  0.1  &     63 & 2.99  &    TRUE \\
      2 &    3  & 0.1   &    63 & 3.03   &   TRUE \\
      2 &   10  & 0.5   &    63 & 3.03   &   TRUE \\
      5 &   10  & 0.1   &    65 & 3.06   &   TRUE \\
        2 &     3  & 0.5  &     63 & 3.06  &     TRUE \\
        5  &  10  & 0.5   &    65 & 3.07   &   TRUE \\
         3  &  10  & 0.1  &     65 & 3.08  &    TRUE \\
         1  &  10  & 0.1  &     63 & 3.15  &    TRUE \\
        1   & 10  & 0.2   &    63 & 3.17   &   TRUE \\
        1   & 10  & 0.5   &    63 & 3.19   &   TRUE \\
        1   &  3  & 0.5   &    63 & 3.23   &   TRUE \\
        5   &  3  & 0.5   &    65 & 3.24   &   TRUE \\
        1   &  3  & 0.2   &    63 & 3.26   &   TRUE \\
         3  &   3 &  0.1  &     64 & 3.30  &    TRUE \\
        3  &  10  & 0.2   &    65 & 3.36   &   TRUE \\
        5   & 10  & 0.2   &    65 & 3.37   &   TRUE \\
        3  &   3  & 0.2   &    65 & 3.40   &   TRUE \\
        3   & 10  & 0.5   &    65 & 3.40   &   TRUE \\
         1  &   3  & 0.1  &     63 & 3.41  &    TRUE \\
        3   &  3  & 0.5   &    64 & 3.47   &   TRUE \\
        5   &  3  & 0.2   &    71 & 3.70   &   TRUE \\
       10   &  3  & 0.2   &   100 & 5.41   &  FALSE \\
       10  &  10  & 0.2   &   100 & 5.47   &  FALSE \\
       10   &  3  & 0.5   &   100 & 5.60   &  FALSE \\
         5   &  3 &  0.1  &    100 & 5.72  &   FALSE \\
        7   & 10  & 0.2   &   100 & 5.87   &  FALSE \\
       10   & 10  & 0.5   &   100 & 5.88   &  FALSE \\
       10   & 10  & 0.1   &   100 & 5.94   &  FALSE \\
        7   &  3  & 0.2   &   100 & 6.00   &  FALSE \\
        7   & 10  & 0.1   &   100 & 6.01   &  FALSE \\
        10  &   3  & 0.1  &    100 & 6.20  &   FALSE \\
         7   &  3  & 0.1  &    100 & 6.25  &   FALSE \\
        7   & 10  & 0.5   &   100 & 6.26   &  FALSE \\
        7  &   3 &  0.5   &   100 & 6.35   &  FALSE \\
    \bottomrule \\
  \end{tabular}  \caption{Number of iterations needed by \inlabru to converge (n iter) considering 100 maximum possible iterations, computational time needed to reach convergence in minutes, and a true/false column reporting if the model converged or not, for different values of  
  parameters of the binning strategy $\delta, \Delta$, and $n_{max}$.}
  \label{tab:5_bins}
\end{table}

\begin{figure}[ht]
    \centering
    \includegraphics[width = 0.8\linewidth, height = 8cm]{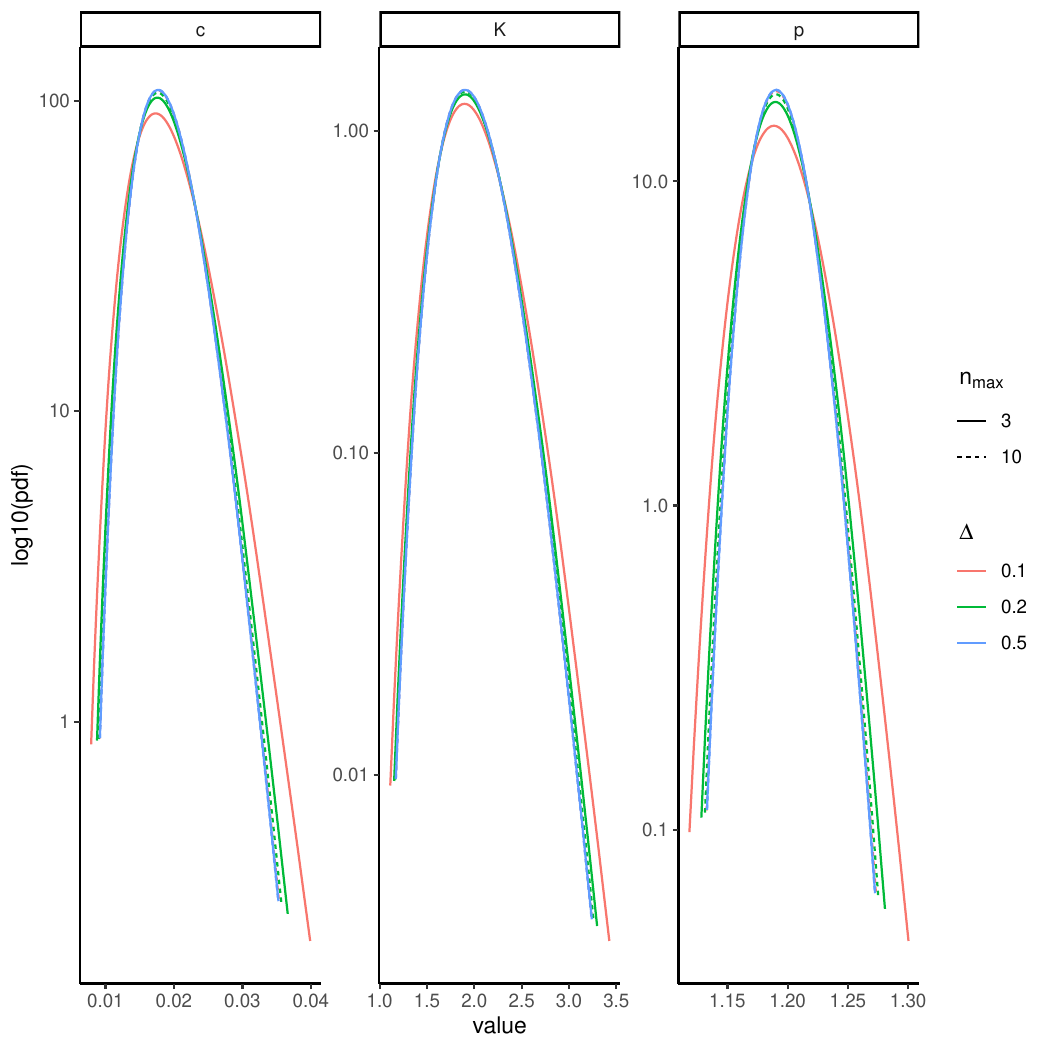}
    \caption{Posterior distribution of ETAS parameters for the \inlabru replicate implementation for different binning strategies. The binning strategies have the same parameter $\delta = 2$, while the others are varying $\Delta = 0.1, 0.2, 0.5$ (color), and $n_{max} = 3, 10$ (line type).}
    \label{fig:10_bin}
\end{figure}

The binning strategy only affects the distribution of the parameters $K, c,$ and $p$: the only parameters of the time triggering function, and therefore, we compare the posterior distributions of these parameters only. We show the posteriors distributions for the case $\delta = 2$ which is the one with the lowest computational time. Figure \ref{fig:10_bin} shows that there are small differences between the models. Only the implementation with $\Delta = 0.1$ and $n_{max} = 3$ has lighter tails, this is due to having too small/not enough bins.   
 
\newpage
\section{Appendix C: Sensitivity to prior choice}
\label{sec:App.C_priorvar}

In this section, we explore the sensitivity of our methodology to change of priors mean and standard deviation. For this task, we chose to use the same prior for all the parameters. We use a Log Gaussian prior with logarithm mean equal to 0 and varying the logarithm standard deviation $\sigma_{log} = 1, 1.5, 2, 2.5$. Table \ref{tab:6_logNsumm} reports summary statistics of the Log Gaussian distribution for the values of $\sigma_{log}$ considered in this analysis.

\begin{table}[!htbp]
    \centering
  \begin{tabular}{llllll}
    \toprule
    $\sigma_{log}$ & mean & sd & q0.025 & q0.5 & q0.975 \\
    \midrule
     1.0 & 1.625  & 2.197   & 0.141 & 1 & 7.099 \\
     1.5 & 3.137  & 8.642   & 0.053 & 1 & 18.915 \\
     2.0 & 6.907  & 43.587  & 0.019 & 1 & 50.397 \\
     2.5 & 28.476 & 144.870 & 0.007 & 1 & 134.278 \\
    \bottomrule \\
  \end{tabular}  \caption{Table reporting summary statistics of Log Gaussian distribution with logarithm mean equal to 0 and logarithm standard deviation $\sigma_{log} = 1, 1.5, 2, 2.5$.}
  \label{tab:6_logNsumm}
\end{table}

\begin{figure}[ht]
    \centering
    \includegraphics[width = 15cm, height = 12cm]{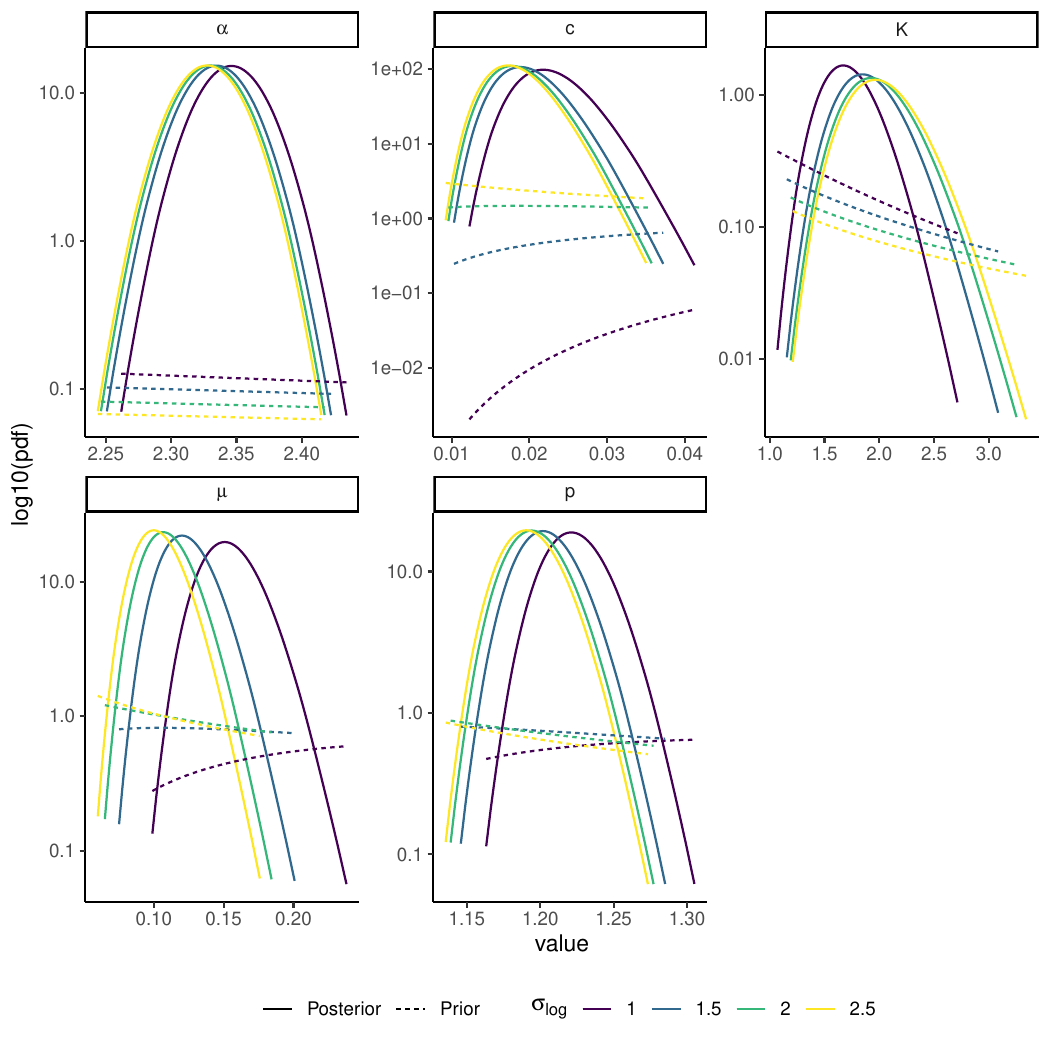}
    \caption{Posterior distribution of ETAS parameters changing the prior mean and standard deviation regulated by the parameter $\sigma_{log}$, the larger the parameter the higher the prior mean and standard deviation. Specifically, we considered $\mu, K, \alpha, c, p - 1 \sim \text{LogN}(0, \sigma_{log})$.}
    \label{fig:11_priorcomp}
\end{figure}

Figure \ref{fig:11_priorcomp} shows that the posterior distributions are robust under the considered changes in prior. Specifically, they appear to converge for increasing values of the prior variance which is what we expect to happen. 

\end{document}